\documentclass[onefignum,onetabnum]{siamart190516}

\usepackage{lipsum}
\usepackage{amsfonts}
\usepackage{graphicx}
\usepackage{epstopdf}

\ifpdf
  \DeclareGraphicsExtensions{.eps,.pdf,.png,.jpg}
\else
  \DeclareGraphicsExtensions{.eps}
\fi

\newsiamremark{remark}{Remark}
\newsiamremark{hypothesis}{Hypothesis}
\crefname{hypothesis}{Hypothesis}{Hypotheses}
\newsiamthm{claim}{Claim}

\headers{Efficient parallelization strategy for real-time FE simulations}{Z. Zeng and H. Courtecuisse}

\title{Efficient parallelization strategy for real-time FE simulations\thanks{Submitted to the editors on May 2 2022.
\funding{This work was supported by French National Research Agency (ANR) within the project SPERRY ANR-18-CE33-0007.}}}

\usepackage{hyperref}
\newcommand{\footremember}[2]{%
    \footnote{#2}
    \newcounter{#1}
    \setcounter{#1}{\value{footnote}}%
}
\newcommand{\footrecall}[1]{%
    \footnotemark[\value{#1}]%
} 

\author{Ziqiu ZENG
\footremember{icube}{ICube, University of Strasbourg, France}
\and Hadrien Courtecuisse
\footrecall{icube} 
\footremember{cnrs}{CNRS, France} }

\usepackage{amsopn}

\ifpdf
\hypersetup{
  pdftitle={Efficient parallelization strategy for real-time FE simulations},
  pdfauthor={Z. Zeng, and H. Courtecuisse}
}
\fi

\usepackage{empheq}
\usepackage{amssymb}
\usepackage{amsmath}
\usepackage{pgfplots}
\pgfplotsset{width=\linewidth, compat=1.9}
\usepackage{multirow}
\usetikzlibrary{patterns}
\usepackage{amsmath,amsfonts,amssymb}
\usepackage{graphicx}
\usepackage{subcaption}
\usepackage{multirow}

\usepackage{mathtools} 
\usepackage{pgfplotstable} 

\pgfplotsset{width=7cm,compat=1.8}

\usepackage{sfmath}
\usepgfplotslibrary{colorbrewer}

\usepackage{ifthen}

\newcommand{\makemath}[1]{{\ensuremath{#1}}}

\newcommand{\dt}{\makemath{h}} 
\newcommand{\timestep}{\makemath{t}} 

\newcommand{\rayleighcoeffmass}{\makemath{\alpha}} 
\newcommand{\rayleighcoeffstiffness}{\makemath{\beta}} 

\newcommand{\pos}{\makemath{\mathbf{ q}}} 
\newcommand{\vel}{\makemath{\dot \pos}} 
\newcommand{\acc}{\makemath{\ddot \pos}} 

\newcommand{\forcevec}{\makemath{\boldsymbol f}} 
\newcommand{\forceextvec}{\makemath{\boldsymbol p}} 

\newcommand{\contactforce}{\makemath{\mathbf{\boldsymbol{\lambda}}}} 
\newcommand{\solutionvec}{\makemath{\boldsymbol{\mathbf{b}}}} 
\newcommand{\unknownvec}{\makemath{\boldsymbol{\mathbf{x}}}} 
\newcommand{\penetration}{\makemath{\boldsymbol{\delta}}} 

\newcommand{\residual}{\makemath{\mathbf{r}}} 
\newcommand{\lowersolvevec}{\makemath{\mathbf{s}}} 

\newcommand{\mass}{\makemath{\mathbf{M}}} 
\newcommand{\damping}{\makemath{\mathbf{B}}} 
\newcommand{\stiffness}{\makemath{\mathbf{K}}} 
\newcommand{\contactJacobian}{\makemath{\mathbf{H}}} 

\newcommand{\systemMat}{\makemath{\mathbf{A}}} 

\newcommand{\delasus}{\makemath{\mathbf{W}}} 

\newcommand{\Lmat}{\makemath{\mathbf{L}}} 
\newcommand{\Dmat}{\makemath{\mathbf{D}}} 

\newcommand{\Accmat}{\makemath{\mathbf{V}}} 

\newcommand{\LmatDiag}{\makemath{\hat{\mathbf{L}}}} 

\newcommand{\LmatAcc}{\makemath{\tilde{{\mathbf{L}}}}} 

\newcommand{\Xmat}{\makemath{\mathbf{X}}} 

\newcommand{\internalforcefunc}{\makemath{\mathcal{F}}} 

\newcommand{\tr}{\makemath{\mathrm{^T}}}

\newcommand{\ldl}{\Lmat\Dmat\Lmat\tr}

\newcommand{\lam}{\makemath{\mathbf{\boldsymbol{\lambda}}}} 
\newcommand{\viol}{\makemath{\mathbf{\boldsymbol{\delta}}}} 

\newcommand{\stiffe}{\makemath{\mathbf{K_\mathit{e}}}} 
\newcommand{\straine}{\makemath{\mathbf{C_\mathit{e}}}} 
\newcommand{\stresse}{\makemath{\mathbf{D_\mathit{e}}}} 

\newcommand{\comments}[3]{\ifthenelse{\equal{1}{#1}}{{#2}}{#3}}

\SetKwProg{Fn}{Function}{}{}
\algnewcommand{\algorithmicand}{\textbf{ and }}
\algnewcommand{\algorithmicor}{\textbf{ or }}
\algnewcommand{\OR}{\algorithmicor}
\algnewcommand{\AND}{\algorithmicand}
\algnewcommand{\var}{\texttt}

\begin{document}

\maketitle

\begin{abstract}
This paper introduces an efficient and generic framework for finite-element simulations under an implicit time integration scheme.
Being compatible with generic constitutive models, a fast matrix assembly method exploits the fact that system matrices are created in a deterministic way as long as the mesh topology remains constant.
Using the sparsity pattern of the assembled system brings about significant optimizations on the assembly stage.
As a result, developed techniques of GPU-based parallelization can be directly applied with the assembled system.
Moreover, an asynchronous Cholesky precondition scheme is used to improve the convergence of the system solver.
On this basis, a GPU-based Cholesky preconditioner is developed, significantly reducing the data transfer between the CPU/GPU during the solving stage.
We evaluate the performance of our method with different mesh elements and hyperelastic models and compare it with typical approaches on the CPU and the GPU.

\end{abstract}  



\section{Introduction}

Medical simulations have received strong interest in providing unlimited access to learn and rehearse complex interventions in a safe environment, without the ethical issues associated. 
Medical simulations have become more and more realistic, offering the possibility to simulate complex interactions in real-time such as contacts and friction between deformable structures.
The general trend is currently focused on the possibility to bring medical simulations closer to the Operating Room (OR) for planning interventions, or even directly in the OR with visual assistance, registration, and augmented reality (AR).
For this purpose, simulations must meet the antagonistic requirements of accuracy and fast computation time at the same time. 
Indeed, advanced finite-elements (FE) simulations are necessary to predict the complex behavior of organs during surgery and provide relevant information for surgeons in real-time.

The material behavior of tissues is generally admitted as non-linear. 
Hyperelastic FE models are nowadays compatible with real-time computations \cite{Marchesseau2010}. 
However, to provide large-scale simulations of detailed meshes, parallelization strategies must be employed to maintain the computational expense sufficiently low to account for user interactions.
In this context, \textit{general-purpose computing on graphics processing units} (GPGPU) has been widely studied because it provides access to massively parallel architecture with very low-cost memory transfers compared to distributed machines. 

Low-level parallelization strategies are necessary to exploit the computational power of the GPUs efficiently.
When applied to FEM, standard approaches aim to accumulate all the elements' contributions in parallel. 
Nevertheless, since several elements share the nodes of meshes, additional operations are necessary to handle concurrent memory access.
Many solutions have been proposed to address this issue.
For instance, \cite{Allard2011} proposed a GPU-based matrix-free approach, enabling high-speed FE simulations with tetrahedral co-rotational elements. 
However, such a technique of parallelizing FE models is very invasive in the code. 
Specific solutions are necessary because each material law leads to different arithmetic operations needed to compute per-element matrices.
Moreover, since memory consumption also depends on the dimension of elements, the ratio between memory access and arithmetic operations brings additional concerns to manage the little cache memory available on the GPUs.
As a result, the algorithm presented in \cite{Allard2011} can hardly be applied in other constitutive models.
A generic parallelization strategy for different models remains a challenge.

On the other hand, as a popular technique used in system solvers, preconditioning boosts convergence and improves performance.
\cite{Courtecuisse2010} introduced an asynchronous preconditioning scheme where a Cholesky preconditioner is factorized in a parallel thread.
This strategy significantly accelerates the convergence since the preconditioner gives a close approximation to the system matrix.
The overhead of factorization is removed from the main simulation loop, making the solving stage very efficient.
However, applying the preconditioner (consisting of solving triangular systems) remains a CPU-based operation, leading to considerable data transfer between the CPU/GPU.
Parallelizing the triangular systems on the GPU remains a challenge, as the forward/backward substitutions lead to data dependency all over the solving stage.

This paper introduces a framework for the system solver of FE simulations.
Based on the currently fastest solving strategy in the \href{www.sofa-framework.org}{Sofa} framework, our main contributions are:
\begin{enumerate}
    \item 
    An efficient matrix assembly method compatible with generic constitutive models is proposed.
    The method requires no specific implementation of the material law on the GPU, but allows efficient solver with typical GPU-based implementation.
    \item 
    A fully GPU-based solving strategy, including the application of the asynchronous preconditioner, is proposed.
    The transfer between the CPU and the GPU is minimal due to an efficient GPU-based Cholesky solver.
\end{enumerate}
These improvements enable efficient GPU-based parallelization for generic constitutive models.

The rest of this paper is organized as follows. 
After reviewing the related works in section \ref{sec: related}, section \ref{sec: Background} presents the relevant deformable models and preconditioning techniques used in this work.
Section \ref{sec: matrix assembly} is dedicated to the fast matrix assembling operation, and section \ref{sec: preconditioner} describes the parallelization strategies. 
In section \ref{sec:interactions} the method is extended to handle collisions and impose boundary conditions. 
Finally, the method is evaluated in section \ref{sec: results} using different FE models.

\section{Related works}
\label{sec: related}

The technical level of computer-based training systems is increasing. Early works in this field proposed simplified models such as mass-spring systems \cite{Kuhnapfel2000}. 
Such discrete methods are simple to implement and fast, but material properties are difficult to parameterize. 
For this reason, they have been progressively replaced by Finite Element (FE) models. 
FE models provide a better understanding of the mechanisms involved in physiological or pathological cases, mainly because the soft-tissue behavior is directly explained through constitutive relations. 
With the rapid growth of computational power, FE models have become compatible with real-time and interactivity. 
First limited to linear elastic models \cite{Bro-Nielsen1996}, it was later extended to large displacements with the co-rotational formulation \cite{Felippa2000}.
FE models are now used for the simulation of hyperelastic or viscoelastic materials in real-time \cite{Marchesseau2010}, with advanced and complex interactions \cite{courtecuisse2014real} between multiple structures.
On the other hand, meshless methods, Position-Based Dynamics (PBD), and Neural Networks are other strategies to model soft tissues in real-time.
A detailed review of this topic goes far beyond the scope of this article, but a survey can be found in \cite{Zhang2018}.

\subsection{Time discretization}

In the context of interactive simulations, an important choice is the time integration scheme.
Indeed, explicit methods have been widely used for medical simulations \cite{Joldes2009}. 
In this case, the solution only involves the (diagonalized) mass matrix leading to very fast, simple to implement, and parallelizable solutions \cite{Comas2008}. 
Unfortunately, user interactions may introduce sudden and stiff contacts at arbitrary location/frequency, which raises stability issues. 

On the opposite, implicit methods are unconditionally stable, i.e., stable (but not necessarily accurate) for any time step and arbitrary stiff materials \cite{LM_Baraff96}. 
Implicit schemes provide better control of the residual vector and hence that the external and internal forces are balanced at the end of the time steps.
Although these advantages come at the cost of solving a set of linear equations at each time step, implicit integration schemes offer a reasonable trade-off between robustness, stability, convergence, and computation time, particularly when combined with a GPU implementation.

\subsection{Solving the set of nonlinear equations}

The nonlinear problem obtained by implicit schemes is usually solved using an iterative Newton Raphson method.
Each iteration of the Newton method consists of solving a linear problem whose solution reduces the error between internal and external forces. 
Since the mechanical matrices depend on the current position (and potentially velocities) of FE meshes, the linear problems must then be recomputed for each simulation step and ideally for each Newton iteration. 
Therefore, the performances of the simulation are directly linked to the efficiency of the solver, which explains why earlier studies mainly focused on sparse linear algebra.
Two families of algorithms are proposed in the literature: direct and iterative methods.

Direct solvers provide the exact solution by computing a factorization (for instance, the Cholesky factorization \cite{Barbic}) or a decomposition (QR decomposition), or eventually, the actual inverse of the system matrix \cite{Bro-Nielsen1996} (though not recommended for large matrices).
As proposed in \cite{George1973}, the nested dissection ordering has been widely used in direct solvers by exploring the parallelism of subproblems while reducing the fill-in of the matrix.
Partitioning and ordering are usually implemented through external tools such as METIS \cite{GeorgeKarypisandVipinKumar}.

The solving phase can then be performed with so-called \textit{forward/backward substitution}, using the two triangular systems (in the case of the Cholesky factorization).
Efficient libraries exist both on the CPU (Pardiso, MUMPS, Taucs) and GPU (cuSPARSE, MAGMA, AmgX). 
The solving stage can be improved by partitioning and reordering the system \cite{Herholz2019}.
Despite the stability of direct solvers, the complete factorization or decomposition of large matrices is usually too time-consuming to be recomputed at each time step, and it is very difficult to parallelize. 
Specific optimizations, inspired by the co-rotational model, have been proposed to incrementally update the sparse Cholesky factorization \cite{Hecht2012} but this approach does not extend to other material laws or element types.

In the interactive context, iterative methods are usually preferred because they limit the number of iterations to compute an approximated solution and better control the time spent during the solving process.
The most popular method is the Conjugate Gradient (CG) algorithm \cite{Saad}, because of the fast convergence and its simple implementation. 
Parallel implementations both on CPU \cite{Parker2009,Hermann2009} and GPU \cite{Bolz,Buatois2009,Allard2011} were proposed.
Nevertheless, the convergence of iterative methods can be significantly impacted for ill-conditioned problems, i.e., when the ratio of the largest and smallest eigenvalues is large. 

\subsection{Matrix assembly and parallelized solver}

The main issue to improve the CG is to gain speedup on sparse matrix-vector multiplication (\textit{SpMV}) operations.
As it is presented in \cite{Bell2009}, to accelerate the \textit{SpMV} operations, many methods are explored to implement them on throughput-oriented processors such as GPU.
Several methods rely on the fact that CG iterations can be performed without explicitly assembling the system matrix \cite{Muller2013,Martinez-Frutos2015}.
Matrix-free methods significantly reduce the memory bandwidth and are proven to be fast and stable. 
However, as a price of speedup, it lacks generality.
As an example, the method introduced in \cite{Allard2011} is designed for the co-rotational formulation and relies on specific cache optimization to compute rotation matrices directly on the GPU. 
However, the specific cache optimizations proposed for the rotation matrices do not extend to other types of material, such as hyperelastic laws.

Explicit assembly of global matrices is necessary for direct solvers to compute the factorization or decomposition of the system. 
The assembly step is usually less critical than the solving process itself, but it may become the bottleneck when combined with efficient solvers. 
There are several ways to construct sparse matrices; the most popular method is first to collect triplets (the row/column index and the value); then compress the triplets in a sparse format. 
A very efficient implementation is provided in the Eigen library. 
Recently \cite{Hiemstra2019} proposed a row by row assembling method for isogeometric linear elasticity problems.
To accelerate the assembling step and minimize memory transfers, several approaches proposed to assemble the matrix directly on the GPU \cite{Dziekonski2012,Zayer,Fu2018}. 
However, specific GPU-based implementation of the assembling procedure is needed for each particular model.

\subsection{Preconditioner}

Another intense area of research aims to improve the performance of the CG algorithm with the use of preconditioners to speed up its convergence.
There are several typical preconditioners: diagonal matrix is simple to build but has limited effect \cite{Baraff1998}; in contrast, precise ones such as incomplete Cholesky factorization are complex and costly to make but can significantly reduce the condition number \cite{Hauth2003}. 

For a typical synchronous preconditioner, the construction of the preconditioner has to be performed before the solving stage of each time integration, leading to additional computation costs.
Some of the recent works aim to find a balance between the cost of applying the preconditioner and the effect of convergence boost, such as efficient preconditioners using the result of incomplete factorization \cite{Anzt2015} and inner Gauss-Seidel preconditioners \cite{Thomas2021}.

On the other hand, the asynchronous preconditioners proposed in \cite{Courtecuisse2010} exploit the continuity of the time line in physically-based simulations.
Relying on the assumption that mechanical matrices undergo relatively small changes between consecutive time steps, the asynchronous preconditioning scheme processes the matrix factorization in a dedicated thread parallel to the main simulation loop and applies the factorization result as a preconditioner after a short delay.
It enables access to a very efficient preconditioner with almost no overhead in the simulation loop.
As a combination of a direct and iterative solver, the method requires explicitly assembling the matrix at a low frequency in the simulation loop to factorize the system in the dedicated thread.
For both synchronous and asynchronous preconditioning schemes, applying the preconditioner requires processing the \textit{forward/backward substitution}, leading to solving sparse triangular systems (STS).
Parallelizing the solution of STS remains challenging in many applications.
There are many works dedicated to improving the performance of STS solvers on the CPU \cite{Bradley2016} and on the GPU \cite{Picciau2017, Yamazaki2020, Li2020b}.
In \cite{courtecuisse2014real}, a GPU-based asynchronous preconditioner was designed to solve the STS with multiple right-hand sides (RHS) in the contact problem.
However, the method cannot efficiently exploit parallelization when dealing with a single RHS.
Therefore, despite the asynchronous preconditioning scheme being introduced with a GPU-based CG implementation of the co-rotational model, applying the preconditioner was performed on the CPU, requiring data transfers between CPU/GPU for each iteration of the preconditioned CG.



\subsection{Implementation}

It is important to note that even each model can be efficiently parallelized on the GPU with specific implementation, it will be hard to be developed and maintained in a large and generic framework such as \textit{SOFA framework}.

A strong motivation of the current work lies in the fact that once the matrices are assembled, the solver can be parallelized independently from the FE models that generated the matrix (i.e., the material law or the type of elements of the FE mesh).
For this purpose, a generic data structure is proposed to fill non-null values of mechanical matrices. 
The method exploits the fact that contributions are added to the system matrix in a deterministic way that only depends on the topology.
Efficient GPU-based parallelization operations such as \textit{SpMV} can be implemented with the assembled system.
A specific parallelization strategy are also proposed to apply the preconditioner on the GPU, allowing this to parallelize the entire solving process independently from the constitutive laws or the type of elements used in the simulation.

\section{Background}
\label{sec: Background}

The current method is based on a general background for deformable simulations using the implicit integration.

\subsection{FE models and constitutive law}

In order to underline the importance of the model, the method is tested with a co-rotational formulation \cite{Morin2017} and with the hyperelastic models \cite{Marchesseau2010}.
The method also applies for various types of elements such as beams, triangles, tetrahedra or hexahedra. 

Using the co-rotational formulation, local stiffness matrices can be precomputed for each element $e$ with the synthetic formulation: 
\begin{equation} 
\stiffe = \int_{V_e} (\straine \stresse \straine^T d V_e)
\label{eqmodel}
\end{equation} 
where $\stresse$ corresponds to the stress-strain matrix, $V_e$ is the volume of the element and $\straine$ is the strain-displacement matrix.
The method is parametrized with the Young modulus $E$ and the Poisson's ratio $\nu$.
The hyperelastic model is implemented with the Multiplicative Jacobian energy decomposition (MJED) method \cite{Marchesseau2010}.
The method consists in decoupling the invariants of the right \textit{Cauchy-deformation tensor} $C = \nabla\varPhi^T \nabla\varPhi$ from the expression of the deformation tensor $I_1 = trC$, $I_2=((trC)^2-trC^2)/2$ and the Jacobian $J= \text{det}\nabla\varPhi$, where $\varPhi$ is the deformation function between the rest and deformed configurations.
The method allows faster stiffness matrix assembly for a large variety of isotropic and anisotropic materials.
In these formulations, the function $\internalforcefunc (\pos, \vel)$ provides internal forces of the deformable body, given the nodal position $\pos$ and velocities $\vel$.

\subsection{Time integration and implicit scheme}

For any time step $\timestep$, the general way to describe the physical behavior of a deformable objective problem can be expressed using Newton's second law:
\begin{equation}
\label{eq:newton's law}
    \mass \acc = \forceextvec - \internalforcefunc (\pos, \vel)
\end{equation}
Where $\mass$ is the mass matrix, $\acc$ the vector of the derivative of the velocity, $\forceextvec$ the external forces and $\internalforcefunc (\pos, \vel)$ the function representing the internal forces.

A backward Euler method is used to integrate the time step.
The implicit scheme can be expressed as follows, where $\dt$ is the length of time interval $[\timestep, \timestep + \dt]$:
\begin{equation}
\label{eq:implicit integration}
    \vel_{\timestep + \dt} = \vel_{\timestep} + \dt \acc_{\timestep + \dt} \quad
    \pos_{\timestep + \dt} = \pos_{\timestep} + \dt \vel_{\timestep + \dt}
\end{equation}

As $\internalforcefunc (\pos, \vel)$ is a non-linear function, a first-order Taylor expansion is performed to linearize the problem \cite{LM_Baraff96}.
This linearization corresponds to the first iteration of Newton-Raphson method. 
The incomplete approximation may cause numerical errors of the dynamic behavior but they lean towards to decrease at equilibrium.

The internal forces are expanded as following:
\begin{equation}
\label{eq:taylor expansion}
    \internalforcefunc ( \pos_{\timestep + \dt} , \vel_{\timestep + \dt} ) = \forcevec_{\timestep} 
    + \frac{\partial \internalforcefunc 
    (\pos , \vel)
    }
    {\partial \pos} 
    \dt \vel_{\timestep + \dt}
    + \frac{\partial \internalforcefunc 
    (\pos , \vel)
    }
    {\partial \vel} 
    \dt \acc_{\timestep + \dt}
\end{equation}
with $\forcevec_{\timestep} = \internalforcefunc (\pos_t, \vel_t)$.

During a time integration, the force function is considered as constant and the partial derivative terms could be expressed as matrices: $\frac{\partial \internalforcefunc}{\partial \vel}$ the damping matrix $\damping$ and $\frac{\partial \internalforcefunc}{\partial \pos}$ the stiffness matrix $\stiffness$. 

By integrating the equations \eqref{eq:newton's law}, \eqref{eq:implicit integration} and \eqref{eq:taylor expansion} we obtain the dynamic equation:
\begin{equation}
\label{eq:dynamic equation1}
    \left(
    \mass + \dt \damping 
    + \dt^2 \stiffness 
    \right)
    \acc_{\timestep + \dt} 
    = \forceextvec_{\timestep} 
    - \left(
    \forcevec_{\timestep}
    + \damping \vel_{\timestep} 
    + \dt \stiffness \vel_{\timestep} 
    \right)
\end{equation}

With Rayleigh damping \cite{LM_Baraff96}, the damping matrix can be expressed as a combination of matrices of mass and stiffness \mbox{$\damping = \rayleighcoeffmass \mass + \rayleighcoeffstiffness \stiffness$} with $\rayleighcoeffmass$ and $\rayleighcoeffstiffness$ the proportional Rayleigh damping coefficients. 
By replacing $\damping$ in the dynamic equation \eqref{eq:dynamic equation1}, it gives:
\begin{equation}
\label{eq:dynamic equation2}
    \underbrace{
    \left[ 
    (1+\dt \rayleighcoeffmass) \mass + \dt (\rayleighcoeffstiffness + \dt) \stiffness 
    \right]
    }_{ \systemMat}
    \underbrace{
    \acc_{\timestep + \dt}
    }_{ \unknownvec}
    =
    \underbrace{
    \forceextvec_{\timestep} 
    - \forcevec_{\timestep}
    + \dt \stiffness
    \vel_{\timestep} 
    }_{ \solutionvec}
\end{equation}

Equation \eqref{eq:dynamic equation2} provides a linear problem $\systemMat \unknownvec = \solutionvec$ to solve.
The left-hand side is a global system matrix $\systemMat$ and the right-hand side a vector $\solutionvec$.
Both of them are constructed by some elements: $\mass$ the mass matrix, $\stiffness$ the stiffness matrix as well as scalar parameters (the time interval and the Rayleigh damping coefficients).
The linear system must be solved at each time step as $\stiffness$ depends on the position of FE models.
Since $\systemMat$ is large and sparse, the general-propose compressed sparse row (CSR) format is used to store the matrix information in three arrays.

\subsection{Asynchronous preconditioner}
\label{sec:async}

Let $\systemMat_{\timestep}$ be the matrix built in a specific time $\timestep$.
Following \cite{Courtecuisse2010}, a preconditioner $\textbf{P}$ can be built from an asynchronous  $\ldl$ factorization:
\begin{equation}
\label{eq:asynchronous preconditioner}
    \textbf{P} = \systemMat_{\timestep} = \ldl
\end{equation}
Where $\Dmat$ is a diagonal matrix and $\Lmat$ a sparse lower triangular matrix. 
The factorized matrices will be available after the factorization is done, normally several time steps after time $\timestep$, and used as a preconditioner with the assumption that \textbf{P} remains a relatively good approximation to the current matrix $\systemMat_{\timestep + n \dt}$. 
In practice, the method is very efficient because the $\ldl$ factorization requires only few simulation steps (usually $n < 5$).

The application of the preconditioner consists mainly of solving the two triangular systems obtained after the factorization. 
The method is very efficient because in practice only $2$ to $5$ preconditioned CG iterations are necessary to converge (with threshold of $10^{-9}$).
However, despite the triangular matrices are sparse and the solution can easily be implemented with a \textit{gauss elimination} on the CPU, this step is difficult to parallelize on a GPU due to numerous data dependencies.
A GPU-based solver with a CPU-based preconditioner leads to considerable data transfer between the different processors, making the solving process inefficient.
Improving this process is an important issue that will be addressed in Section \ref{sec: preconditioner}.

\section{Matrix assembly strategy}
\label{sec: matrix assembly}

Generic constitutive models can benefit from typical GPU-based matrix operations.
But the matrix assembly usually leads to an overhead cost, which is not negligible.
To address the issue, we propose a new assembly approach to meet the requirements of both efficiency and generality.
The fast assembly method relies on the fact that the same assembly procedure is called in each time integration.
When building the system matrix in sequential order, the invariant topology structure brings an important property that the sequence of filling elements into the matrix and the sparsity pattern of $\systemMat$ are definitive.
A specific mapping from the filling element sequence to the final matrix pattern could be built.
As sorting the initial filling sequence to the final sparse format is the most time-consuming stage in the matrix assembly, replacing it with the deterministic mapping brings a significant speedup. 

\begin{figure}[htb!]
    \centering
\includegraphics[width=0.9\linewidth,clip,trim=0 140 0 0]{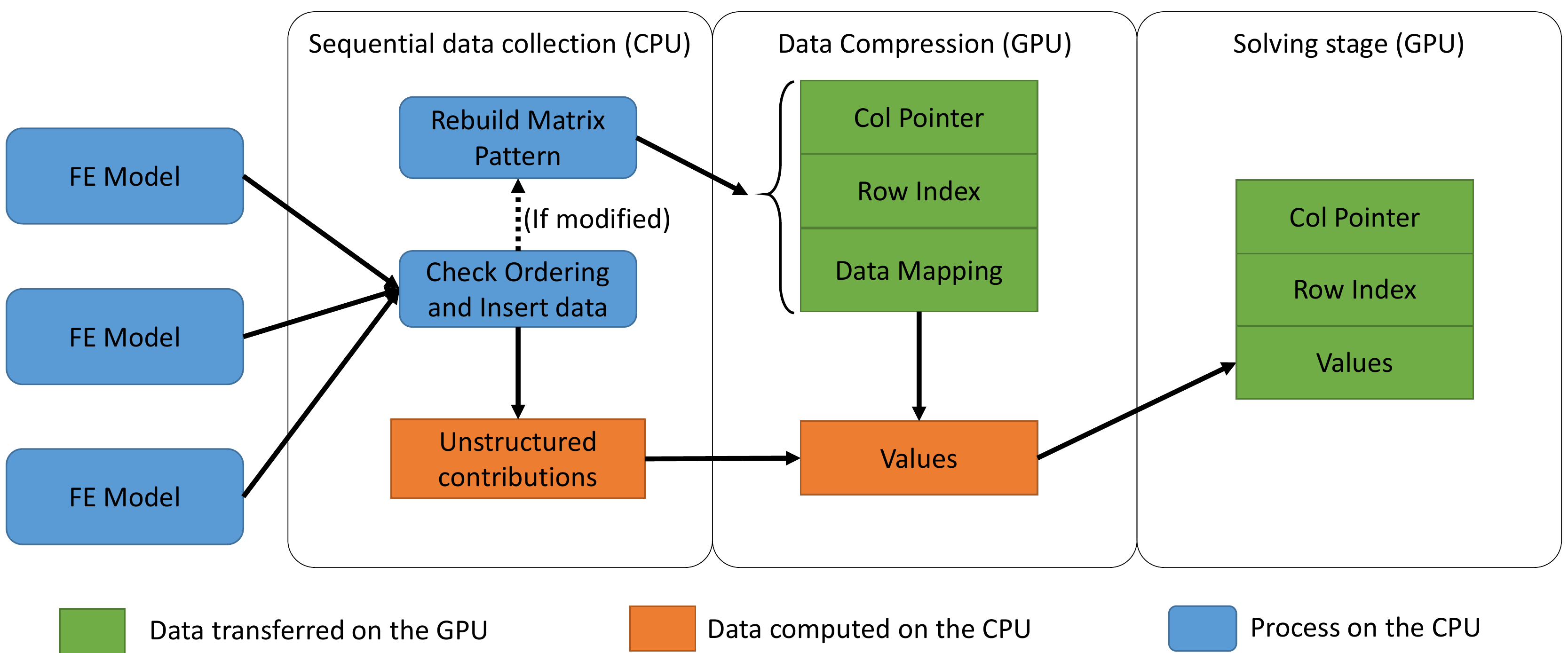}
    \caption{
General workflow of the matrix assembly procedure. 
\textbf{Data Mapping} corresponds to the additional structure used to compress the unstructured contributions (see section \ref{sec:build}) in CSR format.
It is computed and sent on the GPU only once until no modifications of the fill ordering are detected during the collection phase. 
}
    \label{fig:workflow}
\end{figure}

An overview of the general workflow of the assembly procedure is shown in the figure \ref{fig:workflow}.
The matrix assembly consists of the following steps:
\begin{enumerate}
    \item \textbf{Collect data}: Collect the data of mass and stiffness for each element and fill the data in a triplet format (row index, column index, and value).
    \item \textbf{Build matrix pattern}: Sort the collected triplet data by order of row and column. This step is necessary if and only if modifications of the structure have been detected during the collection phase.
    \item \textbf{Compress}: Build the system matrix in CSR format.
\end{enumerate}

The current matrix assembly method relies on the assumption that the topology remains invariant.
Topological modifications are not addressed in this paper but the method remains generic since the topology modification only occurs in specific cases, such as cutting operations.
Applying the asynchronous preconditioning in such case of sudden changes can be addressed with specific correction on the preconditioner \cite{courtecuisse2014real}.
Collisions and interactions may also change the fill ordering of the matrices. 
This specific issue is discussed in section \ref{sec:interactions}.

\subsection{Collect data}
\label{sec:collect}

Matrices $\mass$ and $\stiffness$ in equations \eqref{eq:dynamic equation2} are obtained by summing the local contributions of each element into the global matrices. 
Values are stored in a set of triplets at first, which is a structure containing 3 variables: row index, column index, and value. 
As the triplet vector corresponds to the original process of filling elements into the matrices, the sequence of row/column indices is unsorted and uncompressed\footnote{i.e. the pair row/column may appear several times when filling matrices}, but definitive in each time integration. 
Nevertheless, insertion of contributions into the triplet list will be called many times per second; it must therefore be optimized as much as possible. 
The pseudo-code of the \textbf{add} function is given in the algorithm \ref{algo:add}, and exposed to the FE models in order to insert their contributions.

\SetAlgoInsideSkip{}
\SetAlgoSkip{}
\SetAlFnt{\small}
\begin{algorithm}[htb!]
\SetAlgoLined
\Fn{add(col,row,val)}{
    \uIf{keepStruct \AND id $<$ prevVal.size() \AND prevCol[id] = col \AND  \mbox{prevRow[id] = row}}
    {
        prevVal[id] = val\; 
    }
    \Else 
    {
        keepStruct = \textbf{false}\; 
        prevRow[id] = row \;
        prevCol[id] = col \;
        prevVal[id] = val\;
    }
    id=id+1\;
}
\caption{Procedure used to add value in the matrix.
The boolean \textit{keepStruct} is used to detect any modification in the filling order; \textit{id} gives the next writing address in the uncompressed arrays (\textit{prevRow, prevCol} and \textit{prevVal} corresponding to the triplets list added in the previous time steps).}
\label{algo:add}%
\end{algorithm}

For each inserted value, the test performed line $2$ checks the consistency of the pattern with respect to the previously built matrices. 
This test is a necessary overhead to detect changes in the structure. 
However, if the structure is not modified, only the value \textit{val} is stored (line $7$), allowing this way to take advantage of the cache of the CPU and minimize write operations.

\subsection{Build matrix pattern}
\label{sec:build}

Let $\Xmat$ be a generic matrix to be assembled (such as $\mass$ and $\stiffness$).
A method inspired by the Eigen's library is implemented to build the final CSR format for $\Xmat$.
The method consists of computing twice the transpose of the matrix to sort the values.
To store the temporary matrices, we introduce a temporary format called \textbf{uncompressed structure} which is similar to the CSR format: 
Like the CSR, an arranged row pointer encodes the index in the arrays of column index and values that are unsorted and uncompressed (duplicate indices exist).
We summarize the states of the assembly in different stages in Table \ref{tab:build pattern}.

\begin{enumerate}
\item Firstly, the temporary transposed matrix $\Xmat \tr$ is built in the \textbf{uncompressed structure}.
The computation of the transposed matrix requires beforehand to count the number of values per line, allowing this to allocate the necessary memory. 
Then, data can be moved to their correct location in the allocated structure.
With the pre-defined matrix structure, the sequence of row index can be arranged with a time complexity of $O(2n)$, but inside each row, the sequence of column index remains unsorted.

\item Similar to the previous step, $\Xmat$ is built in the \textbf{uncompressed structure} by transposing $(\Xmat \tr)\tr$.
The second transpose gives the initial matrix $\Xmat$ with a sequence of values sorted both by rows and columns while the structure remains uncompressed.

\item 
Finally, the elements in the same position are merged, transferring the \textbf{uncompressed structure} into the CSR format.
\end{enumerate}

\begin{table}[htb!]
\centering
\begin{tabular}{ |c|c|c|c|c| } 
\hline
Matrix & $\Xmat$ & $\Xmat \tr$  & $\Xmat$ & $\Xmat$  \\
\hline
\hline
Format & triplet set & \shortstack{uncompressed \\ structure} & \shortstack{uncompressed \\ structure} & CSR\\ 
\hline
Row & \shortstack{unsorted \\ uncompressed} & \shortstack{sorted \\ compressed} & \shortstack{sorted \\ compressed} & \shortstack{sorted \\ compressed} \\ 
\hline
\shortstack{Column $\&$ \\ Values} & \shortstack{unsorted \\ uncompressed} & \shortstack{unsorted \\ uncompressed} & \shortstack{sorted \\ uncompressed} & \shortstack{sorted \\ compressed} \\
\hline
\end{tabular}
\caption{State of storage format at different stages in matrix assembly process}
\label{tab:build pattern}
\end{table}

\begin{figure}[t!]
    \centering
    \includegraphics[width=0.8\linewidth]{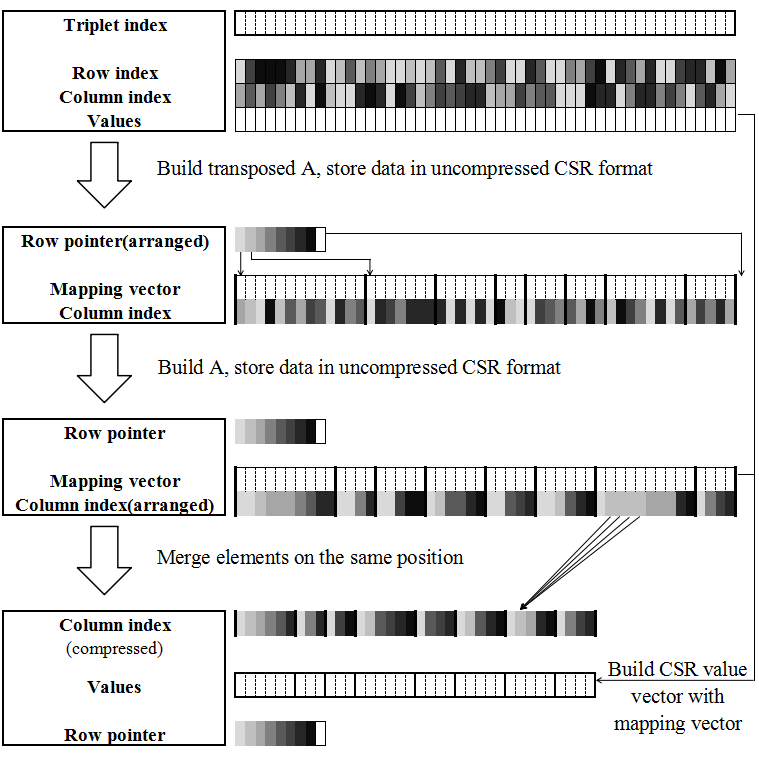}
    \caption{build matrix pattern and mapping vector from triplets set}
    \label{fig:buildpattern}
\end{figure}

One of the main differences with the Eigen's implementation is that the values of the transposed matrices are not directly stored in memory, making a strategy of \textbf{fast assembly} possible:
Relying on the hypothesis that the mesh topology remains unchanged, the filling order, as well as the matrix pattern (row pointer and column index arrays in the CSR), could be reused.
As long as the filling order remains unchanged at the collecting stage, we propose to build a mapping $C$ from the initial triplet set to the CSR format, making it very efficient to build the value array.
Hence, the main operation is to merge the duplicated values in the triplet array that is reordered with the deterministic mapping $C$.
This operation must be performed at each time step, but it can be easily parallelized both on CPU and GPU since the address of values in the CSR format are known and unique. 
More importantly, parallelization can be performed without impacting the code of the constitutive models that generate the matrices (i.e., our method is efficient for any generic constitutive model). 
The deterministic mapping $C$ can be reused as long as no modifications in the filling order are detected at the previous stage. 
On the other hand, in case modifications are detected at the collecting stage, we process the complete rebuilding of the matrix pattern (called \textbf{full assembly}). 
In this mode, the method provides similar performances as the default implementation of Eigen's library.



In order to get the global matrix $\systemMat$ and the vector $\solutionvec$ in equation \eqref{eq:dynamic equation2}, one may note that both are generated from the sum of the same matrices $\mass$ and $\stiffness$ with various coefficients:
\begin{align}
    \systemMat = (1 + \dt \rayleighcoeffmass) \mass + \dt (\dt + \rayleighcoeffstiffness) \stiffness 
    \\
    \solutionvec =
    \forceextvec_{\timestep}  - \forcevec_{\timestep}  
    -
    \dt \stiffness
    \vel_{\timestep}
\end{align}

Coefficients only depend on the time step and the Rayleigh damping constant during the entire simulation. 
Another consequence of our method is that the computation of the right-hand side and the left-hand side terms can be merged in a single procedure, allowing this way to exact a large amount of data that are well suited for GPU architectures, and benefits from cache optimization since the mapping $C$ is accessed twice.

Once the vector of values is compressed, the CSR format can be used directly inside a parallel Conjugate Gradient (either on the CPU or the GPU\footnote{Note that the vector of values is already available on the GPU if the compression is performed on this architecture. The row index and column pointer need to be transferred only if the mapping is modified.}).
For this purpose, the product of the sparse matrix with a vector (\textit{SpMV}) needs to be parallelized, which is trivial with the assembled system. 
Many efficient CPU and GPU-based implementations exist for such a typical operation. 
In this paper, we use the \textit{SpMV} implementation in the CUSPARSE library developed by NVIDIA.
It can significantly improve the time spent in the CG iterations, providing a significant speedup to the entire simulation without modifying the code that generates the matrices. 

\section{System solution}
\label{sec: preconditioner}

The system solution can be efficiently processed with typical sparse matrix operations on GPU with the assembled matrix.
Furthermore, as explained in \ref{sec:async}, the solver can be boosted with a preconditioner.
Compared to the matrix-free method, another important consequence of the fast assembly method lies in the possibility of directly using the assembled matrix to build a preconditioner (which most of the time requires the values of the system explicitly). 
For instance, the diagonal extraction for the Jacobi preconditioner or the lower triangular system for the SSOR preconditioner is exceptionally facilitated.

The asynchronous preconditioner method \cite{Courtecuisse2010} needs to access the explicit values of the assembled system matrix at some specific time steps, which does not add any additional overhead with the proposed assembled solution.
The factorization of the matrix being performed asynchronously, the preconditioner can be entirely computed on the CPU without blocking the main simulation thread.
However, the application of the preconditioner at each iteration of the preconditioned CG implies solving sparse triangular systems. 
The data dependence between lines makes it difficult to be computed in parallel.
Let $\Lmat$ be the lower triangular system of the Cholesky factorization.
We recall the main obstacle for solving a general lower triangular system $\Lmat \lowersolvevec = \residual$ is that the solution $\residual_j$ of a given row $j$ depends on all previous solutions $\lowersolvevec_i$:
\begin{equation}
\label{eq:lower STS}
    \lowersolvevec_j = \residual_j - \sum^{i<j}_{i=0} ( \lowersolvevec_i \Lmat_{j,i} )
\end{equation}
Consequently, the primary operations of the Conjugate Gradient algorithm are processed on the GPU, while the application of the preconditioner remains on the CPU.
This hybrid solving strategy generates a huge amount of data transfer between the processors:
in each CG iteration, the processors need to send the residual vector from GPU to CPU to apply the preconditioner and then send the result vector back to GPU.
In addition to their cost, data transfers impose numerous synchronizations between CPU/GPU, reducing the efficiency of the preconditioner. 
We aim to implement a GPU-based preconditioner that is at least as efficient as the CPU-based version to address the issue.

\subsection{GPU-based Cholesky preconditioner}
We propose a GPU-based Cholesky preconditioner, which is inspired by the solver in \cite{courtecuisse2014real} for sparse triangular systems with multiple right-hand sides.
Since the method in \cite{courtecuisse2014real} was originally designed for multiple RHS. 
When applied in the problem with a single RHS, the level of parallelism for multiple RHS will be left unused.
We propose to fill in this dimension with parallelism imported by the domain decomposition technique using the nested dissection algorithm.
The nested dissection algorithm is used to reduce the filling of the matrix pattern, recursively dividing the mesh into two parts with nearly the same number of vertices while keeping the divider part at a small scale \cite{George1973}.
Consequently, $\Lmat$ is reordered and partitioned into sub-domains with the indices given by the nested dissection algorithm.
The reordering algorithm partitions the graph as follows:
\begin{equation}
    \underbrace{
    \begin{bmatrix}
    \LmatDiag_{a} &  & \\
    & \LmatDiag_{b} & \\
    \Accmat_{a} & \Accmat_{b} & \LmatAcc_{c}
    \end{bmatrix}
    }_{\LmatDiag_{(a, b, c)}}
    \begin{bmatrix}
    \lowersolvevec_{a} \\
    \lowersolvevec_{b} \\
    \lowersolvevec_{c} 
    \end{bmatrix}
    =
    \begin{bmatrix}
    \residual_{a} \\
    \residual_{b} \\
    \residual_{c} 
    \end{bmatrix}
    \label{eq:regional problem}
\end{equation}
where the \textit{diagonal domains} \textbf{a} and \textbf{b} can be solved independently and the reordering algorithm guarantees that the \textit{separator} \textbf{c} (which requires the solution of \textbf{a} and \textbf{b}) is as small as possible.
The partition and reordering are processed recursively on \textit{diagonal domains} \textbf{a} and \textbf{b} until the block size is small enough.

\begin{figure}[htb!]
    \centering
    \includegraphics[width=1.0\linewidth,clip,trim=0 120 0 40]{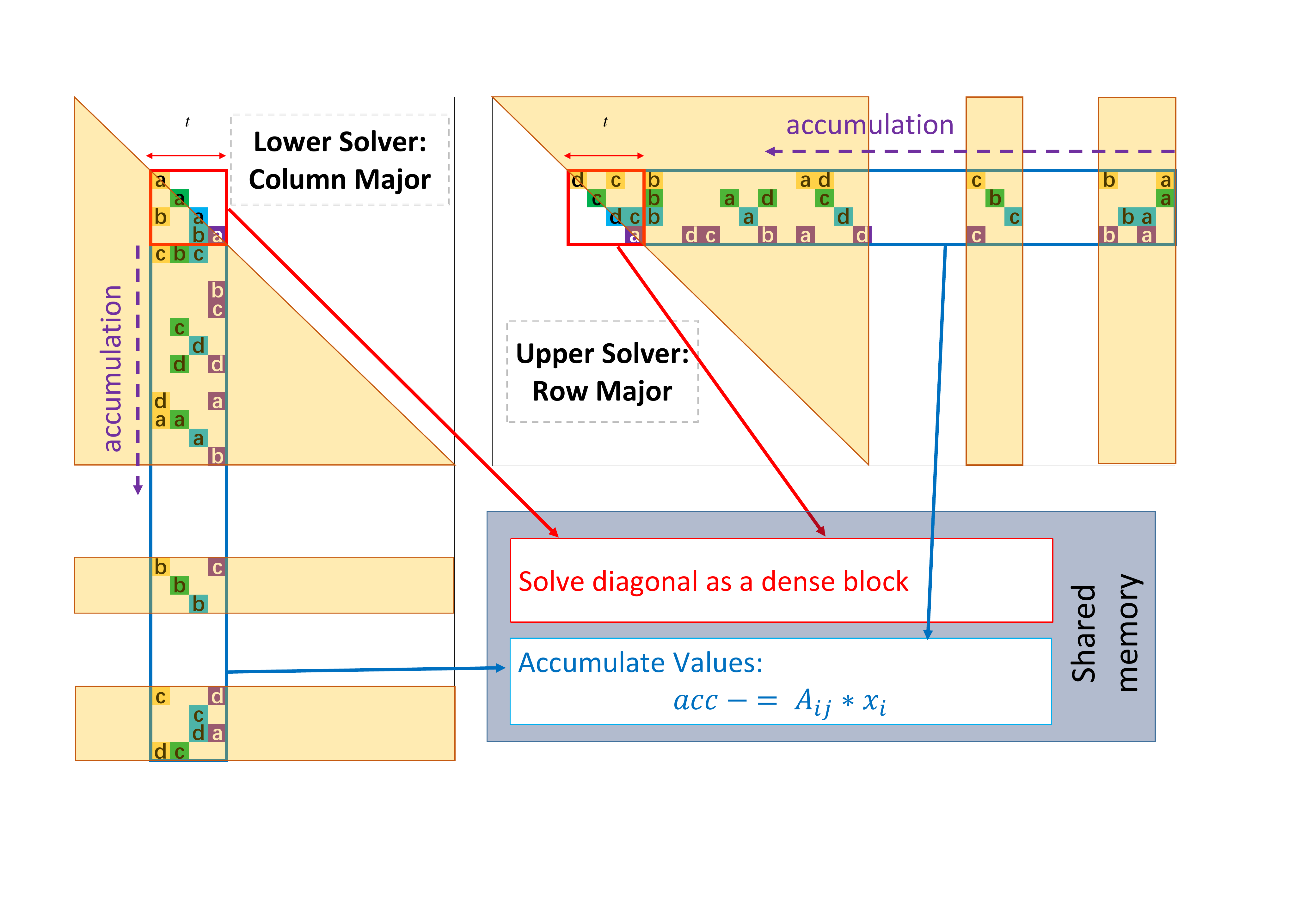}
    \caption{
    The solving stage for each subdomain is realized by GPU kernels, where contributions are accumulated in parallel. 
    For the lower triangular system, the solution can be processed by column sequence (left), which pre-accumulates the data in higher levels, allowing sharing of the computation cost.
    On the other hand, when solving the upper triangular system, computation cost could be shared in lower levels, so the solution needs to be processed oppositely by row sequence (top-right). 
    }
    \label{fig:solveBlock}
\end{figure}

The specific parallelization level is assigned to each subdomain identified in the lower triangular system $\Lmat \lowersolvevec =  \residual$. 
The base rule is that the blocks with higher levels (left edge) require the solutions of low-level blocks.
The upper triangular system problem $\Lmat \tr \lowersolvevec =  \residual$ can be solved with the same method but with an opposite sequence of computation priority. 
The higher level a block has, the less dependence it has.

Within each level, the parallelization strategy presented in \cite{courtecuisse2014real} can be used to solve each \textit{diagonal block} ($\LmatDiag_{a}$, $\LmatDiag_{b}$) and the separator ($\Accmat_{a}$ + $\Accmat_{b}$ + $\LmatAcc_{c}$) by the sequence of rows.
It corresponds to the \textit{Row Major} as illustrated in the figure \ref{fig:solveBlock} where $t*t$ threads are used to accumulate the contribution so that $t$ rows are processed in parallel ($t=16$ in the current implementation). 
Due to the high dependencies, the diagonal part is treated separately as a dense matrix in shared memory.
A parallel reduction is then used to sum the contribution for each row, and finally, the $t*t$ diagonal block is solved as a dense problem.

The opposite \textit{Column Major} is also feasible by pre-accumulating the column's contributions. 
Instead of solving the combined block ($\Accmat_{a}$ + $\Accmat_{b}$ + $\LmatAcc_{c}$)  in a single kernel, the accumulation process of $\Accmat_{a}$ and $\Accmat_{b}$ is moved into the kernel of $\LmatDiag_{a}$ and $\LmatDiag_{b}$ respectively. 
Since it requires only the solution of $\LmatDiag_{a}$ (or $\LmatDiag_{b}$), the accumulation of block $\Accmat_{a}$ (or $\Accmat_{b}$) can be processed in the same kernel. 
The part $\LmatAcc_{c}$ can be solved as a diagonal block after the accumulation of $\Accmat_{a}$ and $\Accmat_{b}$.
Similarly, the diagonal and accumulation parts are treated with $t*t$ threads, and each $t$ column is processed simultaneously. 
This pre-accumulation leads to data writing conflicts since several columns may contribute to the same line simultaneously. 
The atomic add function defined in CUDA can automatically manage the data conflict.

As illustrated in the figure \ref{fig:solveBlock}, in order to share the computation cost in lower levels, the lower solver is implemented with \textit{Column Major}, and the upper system is solved with \textit{Row Major}.
Our level-based parallelization strategy is similar to the approach in \cite{Yamazaki2020}, with several main differences:
\begin{enumerate}
    \item Our solver uses the block-row parallelization strategy in \cite{courtecuisse2014real} to efficiently exploit the parallelism architecture of the GPU (see Figure \ref{fig:solveBlock}).
    \item Our solver is optimized for the problems in FE simulations (e.g. we keep using the analysis result of parallelization level until the matrix pattern is changed).
    \item Our solver benefits from the pre-accumulation technique which allows to share the computation cost in lower levels, making the solver more efficient (see Figure \ref{fig:solveBlock}). 
\end{enumerate}

\subsection{Data Transfer between processors}

\begin{figure*}[htb!]
    \centering
    \includegraphics[width=\linewidth,clip,trim=0 200 0 0]{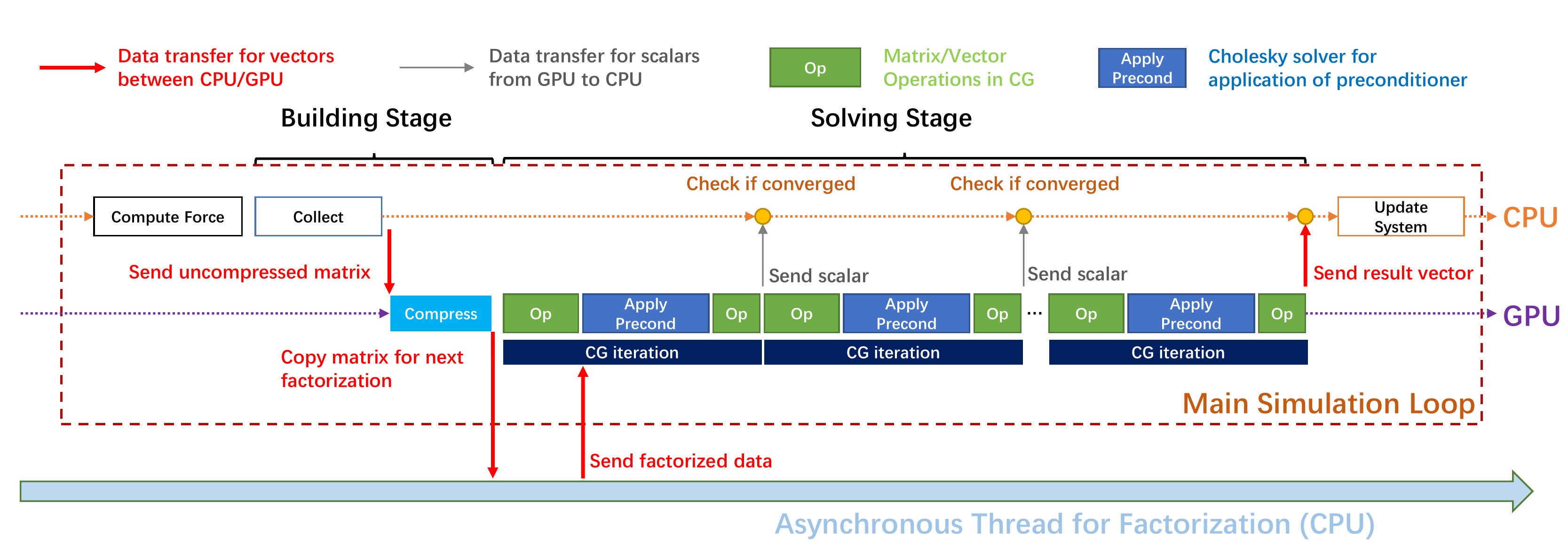}
    \caption{
    Workflow of the asynchronous preconditioning scheme. 
    The \textbf{collect} phase is performed on the CPU with any generic implementation of FE models. 
    The \textbf{Op} process corresponds to the necessary operation to perform one CG iteration. 
    All of them are either \textit{Spmv} or linear algebra operations on vectors that can be easily parallelized on the GPU. 
    The application of preconditioner \textbf{Apply Precond} is also performed on the GPU, resulting in a preconditioned CG fully implemented on GPU.
    Only a scalar needs to be copied to the CPU in each iteration to check the convergence state.
    }
    \label{fig:DataTransfer}
\end{figure*}

We evaluate the performance of our new GPU-based preconditioner in the following section \ref{sec: results}.
As illustrated in the table \ref{tab:precond}, our method is faster than the CPU-based implementation in various examples.
Replacing the CPU-based preconditioner with our GPU-based implementation brings speedup for the solver and addresses the data transfer issue between the processors.
Consequently, our new preconditioner makes it possible to execute a fully GPU-based preconditioned CG, requiring only one scalar to be transferred at each iteration from the GPU to the CPU in order to check the convergence (see Figure \ref{fig:DataTransfer}).

  \section{Contact and interactions}
\label{sec:interactions} 

One can hardly talk about medical simulations without considering interactions between objects. 
The simulation of interacting deformable structure is an extremely large topic; a detailed review can be found in \cite{BOURAGO2005}.

\subsection{Projective Constraints}

The more straightforward solution to fix a set of nodes and impose boundary conditions is to use projective constraints.  
Such constraints consist of reducing the matrix dimension to remove the fixed points from the equations of motion. 
This can be implemented by clearing the corresponding rows and columns of the matrix and setting the value $1$ on the diagonal. 
Clearing a row in the CSR format is trivial, but in order to clear the columns, either a filter could be added in the \textbf{add} procedure of algorithm \ref{algo:add} to skip the undesired values, or a search can be performed on each line afterward in order to erase non-null values on the column if it exists. 
Despite the fact that both strategies are highly time-consuming, they represent the only available alternatives when using the Eigen library. 

With the proposed approach, the undesired values can be identified during the construction of the mapping $C$.
The collection phase is not modified, and all the values associated with fixed points are added in the incoming triplets arrays with no overhead, while indices of fixed rows/columns are stored separately in specific vectors. 
During the computation of the matrix pattern, after the first transpose $\Xmat \tr$, the rows (corresponding to the column of the final matrix) associated with fixed points can be skipped.
Likewise, the undesired lines in the final matrix $(\Xmat \tr)\tr$ can be skipped and replaced by a single value $1$ on the diagonal. 
Finally, the mapping $C$ is built to only assemble the desired values of the projected matrix with no additional overhead in the simulation. 




\subsection{Lagrangian multipliers}

Augmented Lagrangian Multipliers is an efficient solution to deal with constraints accurately and robustly.
The size of the linear systems is increased with specific constraint equations, resulting in a Karush-Kuhn-Tucker (KKT) system:
\begin{equation}
	\label{eq:KKT}
	\begin{cases}
		\systemMat_1 \unknownvec_1 - \contactJacobian_1 \tr \lam =  \solutionvec_1 \\
		\systemMat_2 \unknownvec_2 + \contactJacobian_2 \tr \lam =  \solutionvec_2 \\
		\contactJacobian_1 \unknownvec_1 - \contactJacobian_2 \unknownvec_2 = \Delta \viol
	\end{cases} 
\end{equation}
with subscript $1$ and $2$ representing two interacting objects, $\contactJacobian$ are the linearized constraint equations, $\contactforce$ the associated Lagrangian Multipliers (contact forces) and $\Delta \penetration$ the difference between interpenetration of the end and the beginning of the time step.

Contact constraints can be solved in Linear/Nonlinear Complementary Problem formulations \cite{Duriez2004c}, forming an LCP (linear) to simulate frictionless contact or an NLCP (nonlinear) in case of friction contact \cite{DDKA06}.
The solving process can be performed in several steps: 
\begin{enumerate}
\item \textit{Free motion} The motions are computed without considering the interactions between objects. 
It requires solving the linear systems $\systemMat_1 \unknownvec_1^{free} = \solutionvec_1$ and $\systemMat_2 \unknownvec_2^{free} = \solutionvec_2$.
\item \textit{Constraints resolution} The constraints are defined and the compliance matrix $ \delasus = \contactJacobian_1 \systemMat_1^{-1} \contactJacobian_1^T + \contactJacobian_2 \systemMat_2^{-1} \contactJacobian_2^T$, is built to solve the contact problem $\delasus \contactforce = \delta - \contactJacobian_1 \unknownvec_1^{free} - \contactJacobian_2 \unknownvec_2^{free} $ with the projected Gauss-Seidel.
\item \textit{Motion correction} The motion is corrected solving equations: $\unknownvec_1 =  \unknownvec_1^{free} - \systemMat_1^{-1} \contactJacobian_1^T \contactforce$ and $\unknownvec_2 =  \unknownvec_2^{free} - \systemMat_2^{-1} \contactJacobian_2^T \contactforce$.
\end{enumerate}

A significant advantage of using the Lagrange multipliers is that collision events never modify the system matrix. 
Indeed, the \textit{Free motion} and the \textit{Motion Correction} involve the solution of the same linear system as described in section \ref{sec: matrix assembly} allowing this way to direct benefits for the Fast Assembling method.
The computation of the \textit{compliance matrix} is a time-consuming step that can be significantly improved using the asynchronous preconditioner and the multiple right-hand side solver proposed in \cite{courtecuisse2014real}.
Again, our fast assembly technique significantly improves the construction of the preconditioner resulting in a global speedup of constraint-based simulations.

\section{Results}
\label{sec: results}

\pgfplotsset{every axis/.append style={
                    axis x line=bottom,    
                    axis y line=middle,    
                    axis line style={->}, 
                    label style={font=\tiny},
                    tick label style={font=\tiny},
                    }
                    }

The simulation tests are conducted in the open-source SOFA framework with a CPU Intel@ core i9-9900k at 3.60GHz and a GeForce RTX 2070 8 Gb.



\subsection{Matrix Assembly}
Our matrix assembly strategy aims to reach a compromise between the computation cost and the versatility of the code by assembling the matrix $\systemMat$ with low cost. 
This section compares the matrix building time between the current assembly method and the standard assembly method implemented in the Eigen library. 
The simulation tests for the assembly stage are executed with a group of deformable mesh representing the shape of a raptor with various mesh resolutions (see table \ref{tab:spec}). 

\begin{table}[htb!]
\centering
\begin{tabular}{ |c|c|c|c| } 
\hline
 Example & Raptor 1 & Raptor 2 & Raptor 3 \\
\hline
\hline
Nodes & 2996 & 4104 & 5992\\ 
\hline
Tetra & 8418 & 12580 & 19409 \\ 
\hline
\end{tabular}
\caption{Number of nodes and tetrahedral elements of the meshes.}
\label{tab:spec}
\end{table}


\pgfplotstableread[row sep=\\,col sep=&]{
    Category&Eigen&CPU&GPU\\
        Raptor$\_$1 fast$\_$assembly & 13.06 & 3.55   & 2.38 \\
        Raptor$\_$2 fast$\_$assembly & 18.92 &  5.38   & 3.52 \\
        Raptor$\_$3 fast$\_$assembly & 29.34 & 8.20   & 5.59\\
        Raptor$\_$1 full$\_$assembly  & 12.35 & 12.91   & 18.70 \\
        Raptor$\_$2 full$\_$assembly  & 18.72 & 19.79   & 30.11\\ 
        Raptor$\_$3 full$\_$assembly   & 28.89  & 31.00   & 47.00 \\
}\mydata

\begin{figure}[htb!]
\begin{tikzpicture}
\begin{axis}[
axis x line=middle,    
axis y line=middle,    
axis line style={->}, 
ybar, axis on top,
ylabel={Run time (ms)},
symbolic x coords={
Raptor$\_$1 fast$\_$assembly,
Raptor$\_$2 fast$\_$assembly,
Raptor$\_$3 fast$\_$assembly,
Raptor$\_$1 full$\_$assembly,
Raptor$\_$2 full$\_$assembly,
Raptor$\_$3 full$\_$assembly,
},
x tick label style={align=center, font=\tiny},
xtick=data,
height=8cm,
width=\linewidth,
ymajorgrids, tick align=inside,
major grid style={draw=white},
enlarge y limits={value=.1,upper},
axis x line*=bottom,
axis y line*=left,
bar width=10pt,
ymin=0, ymax=50,
enlarge x limits=.12,
legend style={at={(0.2,0.9)}, anchor=north,legend columns=1,
	font=\tiny},
legend image code/.code={%
       \draw[#1,draw=none,/tikz/.cd,yshift=-0.25em]
        (0cm,1pt) rectangle (6pt,7pt);},
cycle list/Paired,
every axis plot/.append style={fill},
xticklabel style={align=center,text width=12mm},
xtick=data,
nodes near coords={
\pgfmathprintnumber[precision=1]{\pgfplotspointmeta}
},
every node near coord/.append style={font=\tiny},
]
\addplot table[x=Category,y=Eigen]{\mydata};
\addplot table[x=Category,y=CPU]{\mydata};
\addplot table[x=Category,y=GPU]{\mydata};
\legend{Eigen assembly, FA + CPU compression, FA + GPU compression}
\end{axis}
\end{tikzpicture}
\caption{Computational costs for the matrix assembly in \textbf{fast$\_$assembly} mode (operations excluding the re-computation of the compression mapping $C$) and \textbf{full$\_$assembly} mode (when rebuilding the matrix pattern) introduced in Section \ref{sec:build}.
We compare the performance between the Eigen's library implementation (\textbf{Eigen assembly}) and our fast assembly strategy (\textbf{FA}) with the compression performed on the CPU (with 8 CPU threads) or on the GPU.
}
\label{fig:assemlby}
\end{figure}
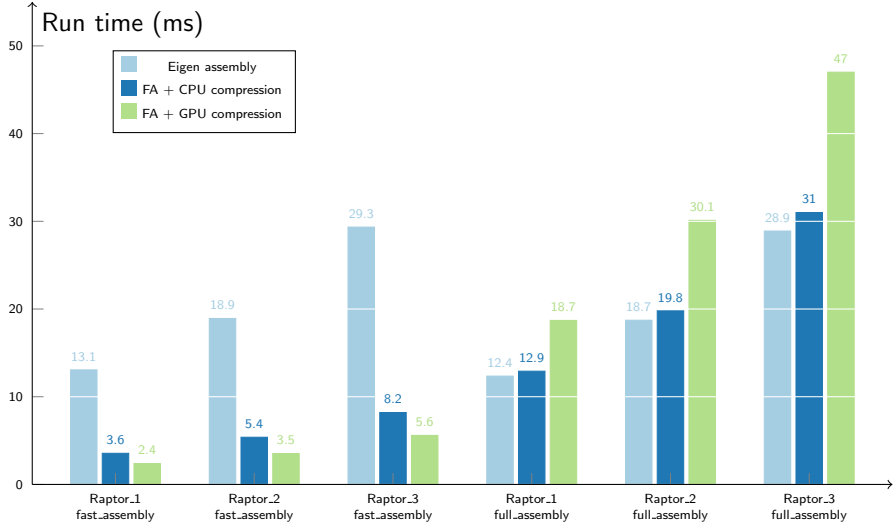
The figure \ref{fig:assemlby} shows the performances of the assembling stage, including the accumulation of triplets and the compression to the CSR format but excluding the computation of the mapping $C$.
With the exception of the first time step where the mapping is actually computed, it corresponds to the standard performances obtained during the entire simulation with the various assembly methods. 
Compared with the standard method using Eigen library, the current method on CPU reduces by $72\%$ time cost of building on average. 
This cost reduction rises to $81\%$ for the fast assembly method on the GPU. 
The compression on the GPU provides a speedup of between $2.7 \times$ to $3 \times$ with respect to the parallel implementation of the compression on the CPU using 8 threads. 

%

If topological modifications are performed or if the filling order is modified, the matrix pattern needs to be rebuilt.
In this case, the building cost, including the computation of the pattern, is measured in the figure \ref{fig:assemlby}.
The time cost of the current method on CPU when the matrix pattern is rebuilt is slightly slower than the Eigen implementation, but it remains in the same order.
The overhead is due to the additional computation of the index vector mapping $C$ providing the position of the triplets in the CSR format.
However, the cost is balanced because the mapping can be reused for the next time steps.
Indeed, reusing the mapping for only two consecutive time steps already provides an acceleration compared to the Eigen implementation. 
Since the computation of the mapping is performed on the CPU, the GPU-based compression suffers a slowdown due to data transfers between the CPU and the GPU.

\subsection{Performances with the CG solver}

\begin{table}[htb!]
\centering
\begin{tabular}{ |c|c|c|c|c| } 
\hline
Example & Liver & Cloth & Cube  & Raptor\\
\hline
\hline
Model  & \multicolumn{3}{c|}{Co-rotational} & Hyperelastic \\
\hline
Type & Tetrahera & Triangle & Hexahedra & Tetrahedra \\
\hline
Nodes & 2660 & 4900 & 8000 & 2996\\ 
\hline
Nb.element & 12328 & 9522& 6859  & 8418 \\ 
\hline
 Mesh &    \begin{minipage}{.1\textwidth}
      \includegraphics[width=\linewidth]{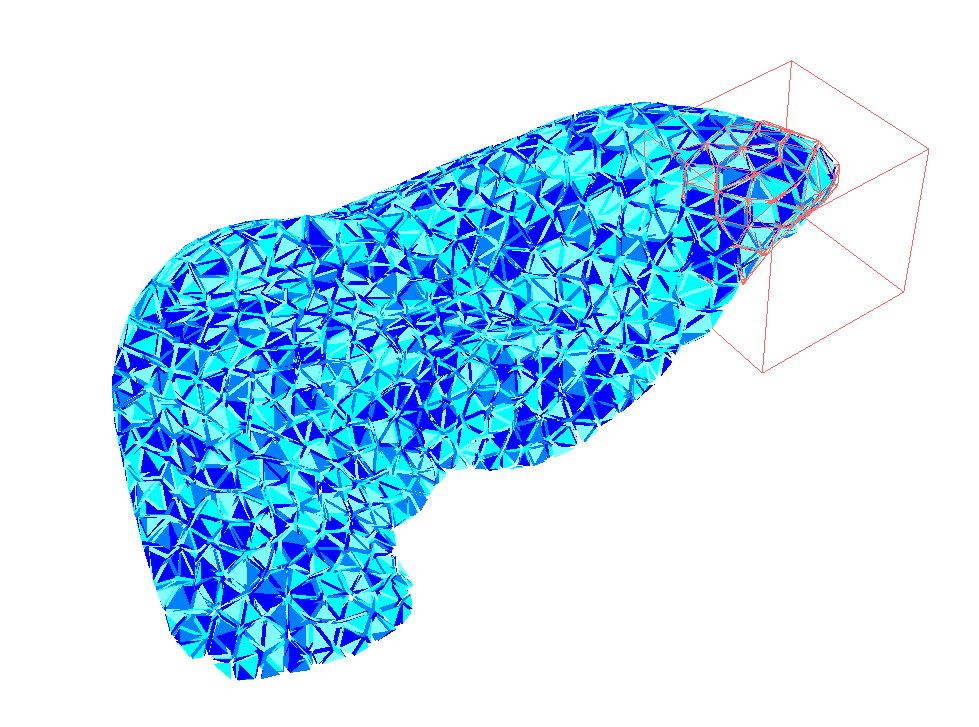}
    \end{minipage}
& 
    \begin{minipage}{.12\textwidth}
      \includegraphics[width=\linewidth]{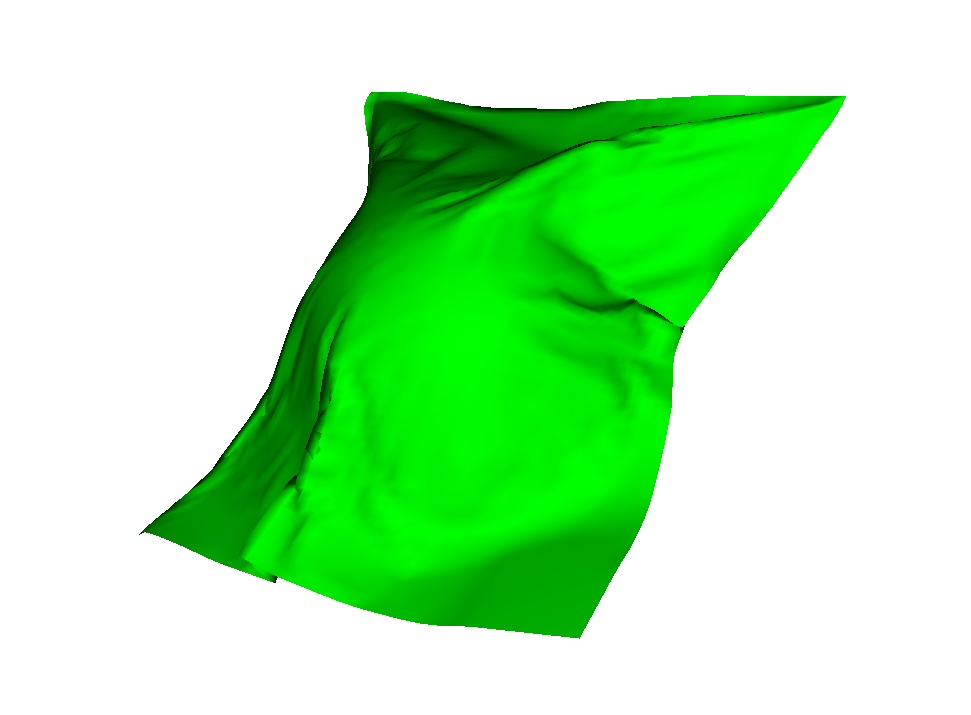}
    \end{minipage}
&
    \begin{minipage}{.1\textwidth}
      \includegraphics[width=\linewidth]{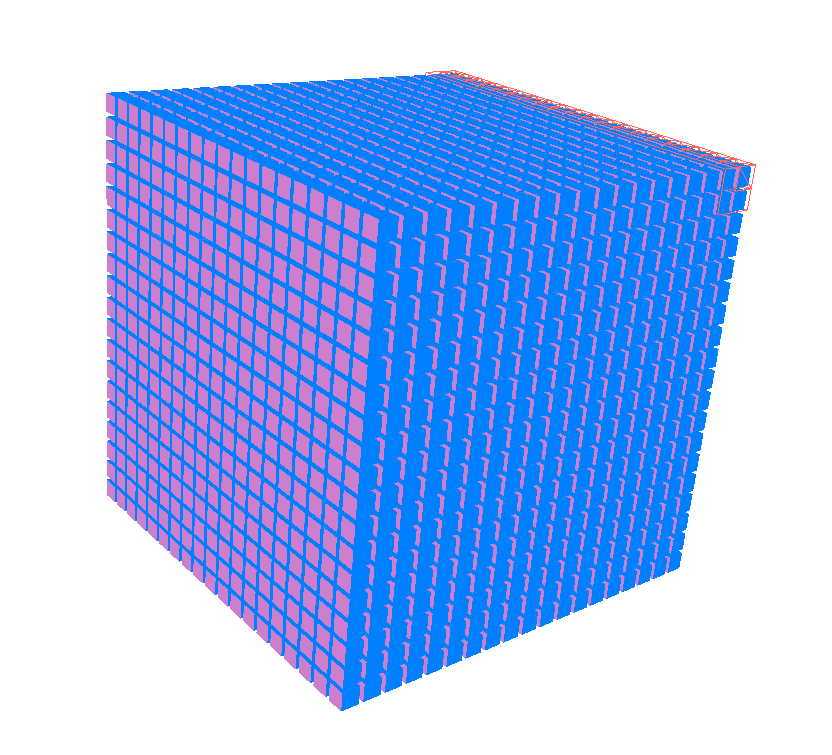}
    \end{minipage}
&
    \begin{minipage}{.15\textwidth}
      \includegraphics[width=\linewidth]{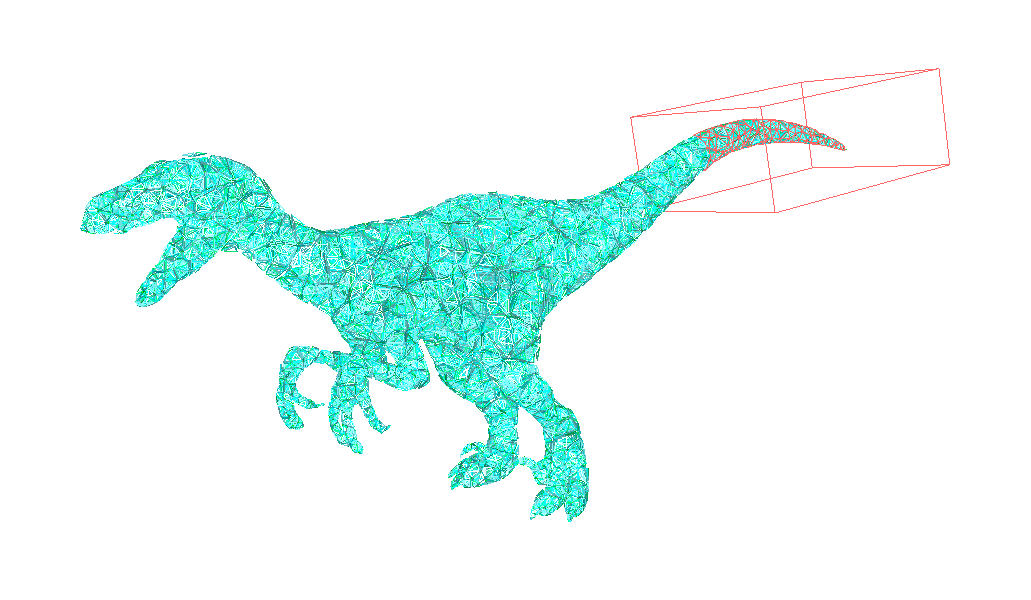}
    \end{minipage}
    \\
\hline
\end{tabular}
\caption{Configurations of different scenario examples.}
\label{tab:specsimu}
\end{table}

The performances of the global simulation are now compared in a complete simulation of a deformable body, including the time for the computation of the FE model, the assembling step and the solving process. 
Performances of the fast assembly method combined with a Conjugate Gradient solver (\textbf{CG GPU fast assembly}) is measured and compared with both a CPU-based matrix-free implementation of the Conjugate Gradient (\textbf{CG CPU matrix-free}) and with the method introduced in \cite{Allard2011} which includes a matrix-free GPU-based Conjugate Gradient (\textbf{CG GPU matrix-free}) for the tetrahedral co-rotational model. 

In order to verify the generality of the proposed solution, the specific GPU-based implementation introduced in \cite{Allard2011} has been extended for other types of elements (triangles and hexahedron), requiring the development of specific code for each model on the GPU.
In addition, the fast assembly method is also tested for hyperelastic material laws.
However, since developing an efficient GPU parallelization is not trivial, the method is only compared with CPU-based matrix-free solvers. 
The scenarios are illustrated in Table \ref{tab:specsimu}.

\begin{figure}[htb!]
\centering%
\noindent%
\begin{subfigure}[t]{1\linewidth}%
\begin{minipage}[t]{1\linewidth}
\begin{tikzpicture}
\begin{axis}[%
    hide axis,
    xmin=-1,
    xmax=1,
    ymin=-1,
    ymax=1,
    legend style={
	legend columns=-1,
	anchor=north west,
	font=\tiny},
    ]
    \addlegendimage{color=blue,mark=square}
    \addlegendentry{CG CPU fast assembly};
    \addlegendimage{color=red,mark=triangle}
    \addlegendentry{CG GPU fast assembly};
    \addlegendimage{color=violet,mark=otimes}
    \addlegendentry{CG CPU matrix-free};
    \addlegendimage{color=black,mark=diamond}
    \addlegendentry{CG GPU matrix-free};
\end{axis}
\end{tikzpicture}
\end{minipage}
\end{subfigure}%
\newline%
\centering%
\noindent%
\begin{subfigure}[t]{0.5\linewidth}%
\begin{minipage}[b]{1\linewidth}
\begin{tikzpicture}
\begin{axis}[
    ymode=log,
    log ticks with fixed point,
    xlabel={CG iterations},
    ylabel={Run time (ms)},
    xmin=0, xmax=1200,
    ymin=5, ymax=800,
    xtick={200,400,600,800,1000},
    ytick={10,20,40,80,160,320,640},
    legend pos=north west,
    ymajorgrids=true,
    grid style=dashed,
]

\addplot[
    color=blue,
    mark=square,
    ]
    coordinates {
    (200,39.20)
(400,66.68)
(600,95.05)
(800,123.75)
(1000,153.18)
    };
\addplot[
    color=red,
    mark=triangle,
    ]
    coordinates {
    (200,15.60)
(400,20.04)
(600,26.10)
(800,32.54)
(1000,38.86)
    };
\addplot[
    color=violet,
    mark=otimes,
    ]
    coordinates {
    (200,130.17)
(400,249.15)
(600,367.80)
(800,511.56)
(1000,621.31)	
    };
\addplot[
    color=black,
    mark=diamond,
    ]
    coordinates {
    (200,8.97)
(400,21.98)
(600,27.90)
(800,36.60)
(1000,48.39)	
    };
\end{axis}
\end{tikzpicture}
\end{minipage}
\caption{Liver mesh (tetrahedron elements)}
\label{fig: CG Tetra Corotational model}
\end{subfigure}%
\begin{subfigure}[t]{0.5\linewidth}%
\begin{minipage}[b]{1\linewidth}
\begin{tikzpicture}
\begin{axis}[
    ymode=log,
    log ticks with fixed point,
    xlabel={CG iterations},
    ylabel={Run time (ms)},
    xmin=0, xmax=1200,
    ymin=5, ymax=400,
    xtick={200,400,600,800,1000},
    ytick={10,20,40,80,160,320},
    legend pos=north west,
    ymajorgrids=true,
    grid style=dashed,
]

\addplot[
    color=blue,
    mark=square,
    ]
    coordinates {
    (200,40.63)
(400,75.68)
(600,109.62)
(800,145.89)
(1000,182.40)
    };
\addplot[
    color=red,
    mark=triangle,
    ]
    coordinates {
    (200,12.90)
(400,18.06)
(600,26.81)
(800,32.25)
(1000,44.53)
    };
\addplot[
    color=violet,
    mark=otimes,
    ]
    coordinates {
    (200,36.76)
(400,72.60)
(600,108.54)
(800,144.24)
(1000,177.94)	
    };
\addplot[
    color=black,
    mark=diamond,
    ]
    coordinates {
    (200,8.66)
(400,17.30)
(600,27.62)
(800,40.16)
(1000,49.65)	
    };
\end{axis}
\end{tikzpicture}
\end{minipage}
\caption{Cloth mesh (Triangular elements)}
\label{fig: CG Triangle linear model}
\end{subfigure}%
\newline%
\centering%
\noindent%
\begin{subfigure}[t]{0.5\linewidth}%
\begin{minipage}[b]{1\linewidth}
\begin{tikzpicture}
\begin{axis}[
    ymode=log,
    log ticks with fixed point,
    xlabel={CG iterations},
    ylabel={Run time (ms)},
    xmin=0, xmax=600,
    ymin=5, ymax=2560,
    xtick={100,200,300,400,500},
    ytick={10,20,40,80,160,320,640,1280},
    legend pos=north west,
    ymajorgrids=true,
    grid style=dashed,
]
\addplot[
    color=blue,
    mark=square,
    ]
    coordinates {
    (100,40.63)
(200, 75.68)
(300,109.62)
(400,145.89)
(500,182.40)
    };
\addplot[
    color=red,
    mark=triangle,
    ]
    coordinates {
    (100,37.87)
(200,48.65)
(300,59.18)
(400,63.31)
(500,75.25)
    };
\addplot[
    color=violet,
    mark=otimes,
    ]
    coordinates {
    (100,286.67)
(200,561.09)
(300,838.50)
(400,1112.96)
(500,1381.29)	
    };
\addplot[
    color=black,
    mark=diamond,
    ]
    coordinates {
    (100,21.32)
(200,42.12)
(300,62.33)
(400,84.31)
(500,103.62)	
    };
\end{axis}
\end{tikzpicture}
\end{minipage}
\caption{Cube mesh (hexahedron elements)}
\label{fig: CG Hexa linear model}
\end{subfigure}%
\caption{Computation time of a single time step for different examples modeled with the co-rotational formulation for various (fixed) number of CG iterations.
The figures share the same legend on the top. The details of different examples can be found in Table \ref{tab:specsimu}. We note that the y axis is logarithmic in these figures.
}
\label{fig:co-roatation}
\end{figure}
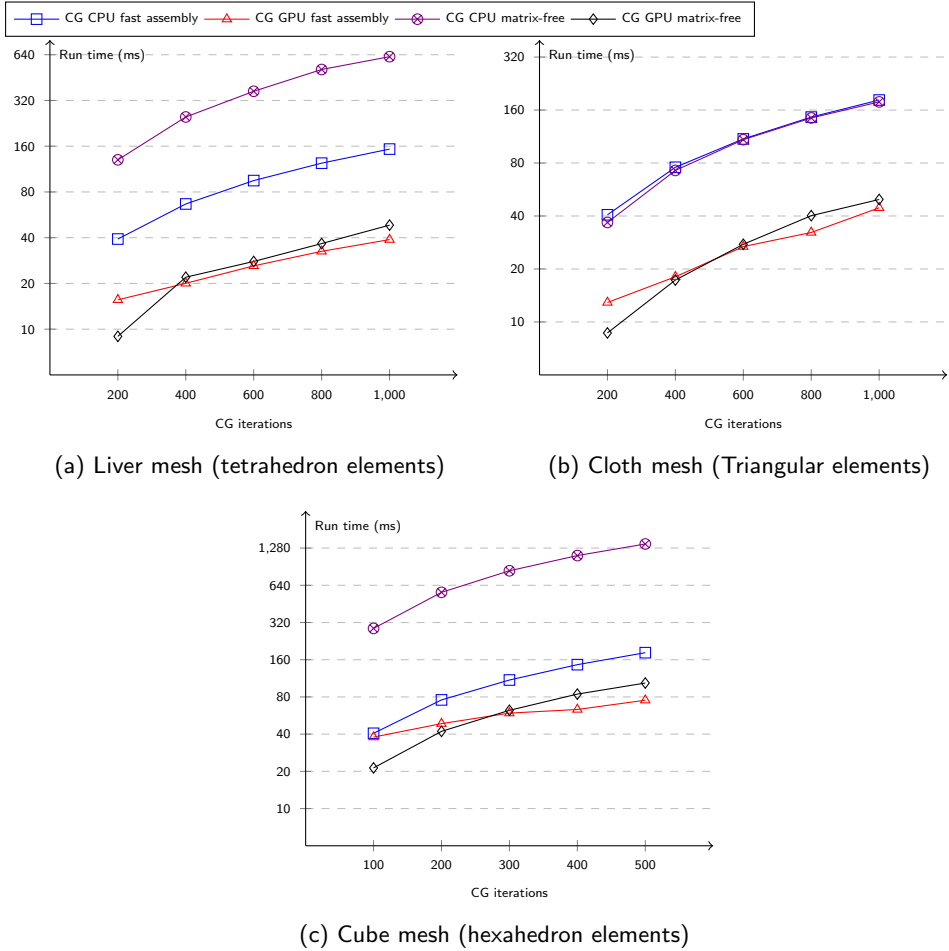

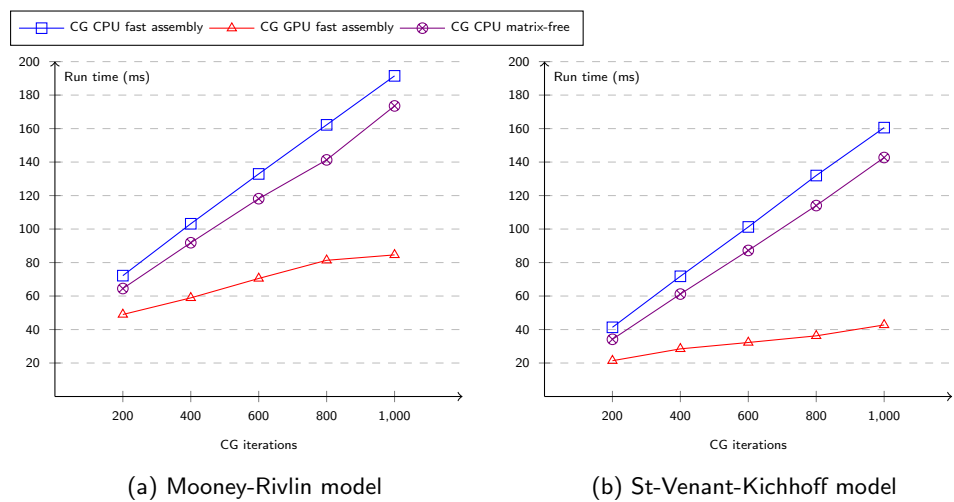
\begin{figure}[htb!]
\centering%
\noindent%
\begin{subfigure}[t]{1\linewidth}%
\begin{minipage}[t]{1\linewidth}
\begin{tikzpicture}
\begin{axis}[%
    hide axis,
    xmin=-1,
    xmax=1,
    ymin=-1,
    ymax=1,
    legend style={
	legend columns=-1,
	anchor=north west,
	font=\tiny},
    ]
    \addlegendimage{color=blue,mark=square}
    \addlegendentry{CG CPU fast assembly};
    \addlegendimage{color=red,mark=triangle}
    \addlegendentry{CG GPU fast assembly};
    \addlegendimage{color=violet,mark=otimes}
    \addlegendentry{CG CPU matrix-free};
\end{axis}
\end{tikzpicture}
\end{minipage}
\end{subfigure}%
\newline%
\centering%
\noindent%
\begin{subfigure}[t]{0.5\linewidth}%
\begin{minipage}[b]{1\linewidth}
\begin{tikzpicture}
\begin{axis}[
    xlabel={CG iterations},
    ylabel={Run time (ms)},
    xmin=0, xmax=1200,
    ymin=0, ymax=200,
    xtick={200,400,600,800,1000},
    ytick={0,20,40,60,80,100,120,140,160,180,200},
    legend pos=north west,
    ymajorgrids=true,
    grid style=dashed,
	legend style={at={(0.7,0.98)},
	anchor=north,legend columns=1,
	font=\tiny},
]

\addplot[
    color=blue,
    mark=square,
    ]
    coordinates {
    (200,72.18)
(400,103.16)
(600,132.92)
(800,162.23)
(1000,191.48)
    };
\addplot[
    color=red,
    mark=triangle,
    ]
    coordinates {
    (200,48.98)
(400,58.91)
(600,70.50)
(800,81.34)
(1000,84.57)
    };
\addplot[
    color=violet,
    mark=otimes,
    ]
    coordinates {
    (200,64.48)
(400,91.83)
(600,118.13)
(800,141.30)
(1000,173.53)	
    };
\end{axis}
\end{tikzpicture}
\end{minipage}
\caption{Mooney-Rivlin model}
\label{fig: CG tetra Mooney-Rivlin model}
\end{subfigure}%
\begin{subfigure}[t]{0.5\linewidth}%
\begin{minipage}[b]{1\linewidth}
\begin{tikzpicture}
\begin{axis}[
    xlabel={CG iterations},
    ylabel={Run time (ms)},
    xmin=0, xmax=1200,
    ymin=0, ymax=200,
    xtick={200,400,600,800,1000},
    ytick={0,20,40,60,80,100,120,140,160,180,200},
    legend pos=north west,
    ymajorgrids=true,
    grid style=dashed,
	legend style={at={(0.7,0.98)},
	anchor=north,legend columns=1,
	font=\tiny},
]

\addplot[
    color=blue,
    mark=square,
    ]
    coordinates {
    (200,41.32)
(400,71.79)
(600,101.26)
(800,131.96)
(1000,160.57)
    };
\addplot[
    color=red,
    mark=triangle,
    ]
    coordinates {
    (200,21.38)
(400,28.46)
(600,32.26)
(800,36.23)
(1000,42.74)
    };
\addplot[
    color=violet,
    mark=otimes,
    ]
    coordinates {
    (200,34.16)
(400,61.18)
(600,87.22)
(800,114.03)
(1000,142.72)	
    };
\end{axis}
\end{tikzpicture}
\end{minipage}
\caption{St-Venant-Kichhoff model}
\label{fig: CG tetra st venant model}
\end{subfigure}%

\caption{Computation time of a single simulation step with hyperelastic models implemented in SOFA for various (fixed) number of CG iterations per time step. The figures share the same legend on the top. The scenarios simulate a tetrahedron raptor mesh (see Table \ref{tab:specsimu}).}%
\label{fig:hyperelastic}
\end{figure}


For the scenarios in Figure \ref{fig:co-roatation}, the run time increases linearly along with the number of iterations.
The fast assembly method combined with the GPU-based CG is up to $16 \times$ faster than the standard CPU method implemented in SOFA and reaches the same computation cost level as the GPU-optimized method.
The Fast Assembly method suffers a slowdown compared to the GPU matrix-free method with fewer iterations, but this case is inverted when the iteration increases. 
This is due to the fact that the fast assembly method takes time to build the matrix, but this overhead is compensated at each CG iteration since the parallel implementation of the \textit{SpMV} operation is faster with the assembled matrix.

It's important to note that although performances are comparable to the GPU-based matrix-free implementation, the code of the co-rotational model is written for the CPU where optimizations are simply obtained by calling the \textbf{add} function of the algorithm \ref{algo:add}, which is completely transparent for the code and enforces the compatibility with the rest of the models implemented in the SOFA framework. 
In addition, for computers without GPU-compatible hardware, the \textit{SpMV} operation can also be parallelized on the CPU.
The method \textbf{CG CPU fast assembly} uses 8 threads to perform the matrix-vector product, which leads to a speedup of up to $4.13 \times$ compared to the sequential method (see the figure \ref{fig: CG Tetra Corotational model}).


In the figure \ref{fig: CG tetra Mooney-Rivlin model}, the method is directly tested with the Mooney-Rivlin material using the implementation of the MJED \cite{Marchesseau2010} provided in SOFA, without any modification of the code. 
The main difference with the co-rotational formulation lies in the fact that the computation of the hyperelastic formulation is significantly slower, and thus the time spent in the assembling and solving processes is smaller. 
Therefore, the benefits of the CPU parallelization with 8 threads (\textbf{CG CPU fast assembly}) is balanced by the overhead of assembling the matrix compared to the matrix-free version (\textbf{CG CPU matrix-free}). 
However, the GPU-based internal parallelization of the assembling and solving process provides a speedup between $1.31\times$ and $2.05\times$.
This represents the fastest method for nonlinear materials available in SOFA because no specific GPU-based parallelization of the MJED method is available. 
The method is also tested with the St Venant-Kirchhoff model using the MJED implementation.
Compared to the Mooney-Rivlin material, the model is less complex so that the computation of the hyperelastic formulation is less costly.
In the figure \ref{fig: CG tetra st venant model}, the fast assembly method gains a speedup between $1.60\times$ and $3.34\times$ compared to the matrix-free method, which is the fastest current implementation for the nonlinear model in SOFA.

\subsection{Performances with the preconditioned CG solver}

The performances of different sparse $\ldl$ solvers (including both the lower and upper triangular systems) are reported in the table \ref{tab:perfsts}.
The proposed GPU-based parallelization relying on the nested dissection method (\textbf{GPU ND}) introduced in section \ref{sec: preconditioner} is $20.3 - 24.0 \times$ faster than the GPU-based implementation provided in NVIDIA's CUSPARSE library.
The main reasons lie in the fact that the CUSPARSE method requires performing the analysis of the data dependencies before actually solving the problem, and the parallelization strategies are optimal for much larger problems than the ones used in the context of real-time simulations. 
Such speedup compared to the golden-standard implementation (CUSPARSE library) is reported as maximally $5.8 \times$ in \cite{Picciau2017} and $19.5 \times$ in \cite{Yamazaki2020}.

The method is also compared with the GPU-based implementation proposed in \cite{courtecuisse2014real}.
As reported in this previous work, the GPU-based $\ldl$ solver is $3\times$ slower than a sequential CPU implementation, whereas the (\textbf{GPU ND}) provides a speedup of $1.4 - 2 \times$, enabling the possibility to solve the problem directly on the GPU. 

\begin{table}[htb!]
\centering
\begin{tabular}{ |c|c|c|c|c| } 
\hline
Mesh & Method & LDL solver & Lower & Upper \\
\hline
\hline
\multirow{4}{4em}{Raptor 1} 
& CUSPARSE & 13.46 & 3.33 & 2.48\\ 
& \cite{courtecuisse2014real} GPU & 3.63 & 1.88 & 1.66\\ 
& CPU &1.13 & 0.52 & 0.58 \\ 
& GPU ND & 0.56 & 0.29 & 0.24 \\ 
\hline
\multirow{4}{4em}{Raptor 2} 
& CUSPARSE & 22.77 & 5.63 & 3.91\\ 
& \cite{courtecuisse2014real} GPU & 6.36 & 3.18 & 2.78\\ 
& CPU & 1.97 & 0.90 & 1.04 \\ 
& GPU ND & 1.12 & 0.60 & 0.49\\ 
\hline
\multirow{4}{4em}{Raptor 3} 
& CUSPARSE & 44.96 & 11.02 & 6.97\\ 
& \cite{courtecuisse2014real} GPU & 10.07 & 5.33 & 4.71\\ 
& CPU & 3.94 & 1.77 & 2.14\\ 
& GPU ND & 2.15 & 1.07 & 1.05 \\ 
\hline
\end{tabular}
\caption{Computation time (in ms) of various STS solvers}
\label{tab:perfsts}
\end{table}

\begin{table}[htb!]
\setlength{\tabcolsep}{4pt}
\centering
\begin{tabular}{ |c|c||c|c|c||c|c|c| } 
\hline
 & \multirow{2}{*}{$\#$it} & \multicolumn{3}{c||}{Method} & \multirow{2}{*}{Raptor 1} & \multirow{2}{*}{Raptor 2} & \multirow{2}{*}{Raptor 3} \\
 &  & Assembly & CG & Precond &  &  &  \\
\hline
\hline
\multirow{4}{*}{\rotatebox{90}{Corot}} & \multirow{4}{*}{15} & Fast Assembly & CPU & CPU & 25.94 & 45.85 & 85.86 \\ 
\cline{3-8}
& & \textbf{Fast Assembly} & \textbf{GPU} & \textbf{GPU} & \textbf{14.58} & \textbf{26.46} & \textbf{46.84} \\ 
\cline{3-8}
& & Matrix Free & CPU & CPU & 43.18 & 65.65 & 118.33 \\ 
\cline{3-8}
& & Matrix Free & GPU & CPU & 31.00 & 48.52 & 80.50 \\ 
\hline \hline

\multirow{3}{*}{\rotatebox{90}{\small MR}} & \multirow{3}{*}{12} & Fast Assembly&  CPU & CPU &59.02 & 90.64 & 153.90\\ 
\cline{3-8}
& & \textbf{Fast Assembly} & \textbf{GPU} & \textbf{GPU} & \textbf{48.73} & \textbf{77.84} & \textbf{124.53} \\ 
\cline{3-8}
& &  Matrix Free & CPU & CPU & 56.76 & 89.76 & 152.11 \\ 
\hline \hline

\multirow{3}{*}{\rotatebox{90}{\small SVK}} & \multirow{3}{*}{8} & Fast Assembly & CPU &  CPU & 23.16 & 37.96 & 67.03 \\ 
\cline{3-8}
& & \textbf{Fast Assembly} & \textbf{GPU} & \textbf{GPU} & \textbf{16.60} & \textbf{26.46} & \textbf{42.86} \\ 
\cline{3-8}
& &  Matrix Free & CPU & CPU & 24.18 & 37.35 & 64.67 \\ 
\hline
\end{tabular}
\caption{Computation time (in ms) for various models: Corotational (\textbf{Corot}), Mooney-Rivelin (\textbf{MR}) and St-Venant-Kichhoff (\textbf{SVK}).}
\label{tab:precond}
\end{table}

The method is tested in complete simulations of deformable bodies with various constitutive laws (see Table \ref{tab:precond}).
The tests are conducted with the same mesh group of raptors and solved with the asynchronous preconditioned CG.
On average, the asynchronous preconditioner is updated every $2$ to $4$ simulation steps, which lead between $5$ to $20$ iterations (\textbf{$\#$it}) according to different cases.
Therefore, the asynchronous preconditioner already provides a significant speedup with respect to the standard GPU-based CG.
With the preconditioner, the matrix operations needed during the CG iterations are performed either with the fast assembly method or with the matrix-free method , either on the CPU or the GPU when available. 
The preconditioner is explicitly built using the fast assembly method and applied on the CPU as done in \cite{courtecuisse2014real} or on the GPU with the method introduced in section \ref{sec: preconditioner}.

The method \textbf{fast assembly + CG GPU + preconditioner GPU} is the fastest method and provides a speedup of between $1.7\times$ and $2.1\times$ for the co-rotational model compared to the solution proposed in \cite{courtecuisse2014real}.
An important advantage of the current solution is that the preconditioned CG is applied entirely on the GPU, without any need for data transfers or synchronizations between the CPU/GPU during the solving stage.
In addition, since the matrix is assembled every time step, no additional overhead is introduced when the factorization needs to be recomputed.

As no GPU-based matrix-free method is implemented for hyperelastic models in SOFA, the comparison is made with the CG performed on the CPU.
Although the computation cost of the hyperelastic formulation is significantly higher, the result of the proposed GPU version also provides a speedup from $1.15 \times$ to $1.22 \times$ for the Mooney-Rivlin material.
For the St Venant-Kirchhoff material, where the model is more straightforward than the Mooney-Rivlin material, this speedup is raised from $1.41 \times$ to $1.51 \times$.

\subsection{Contact and interactions}

\begin{figure}[htb!]
\centering%
\noindent%
\begin{subfigure}[t]{0.3\linewidth}%
\includegraphics[width=.9\linewidth]{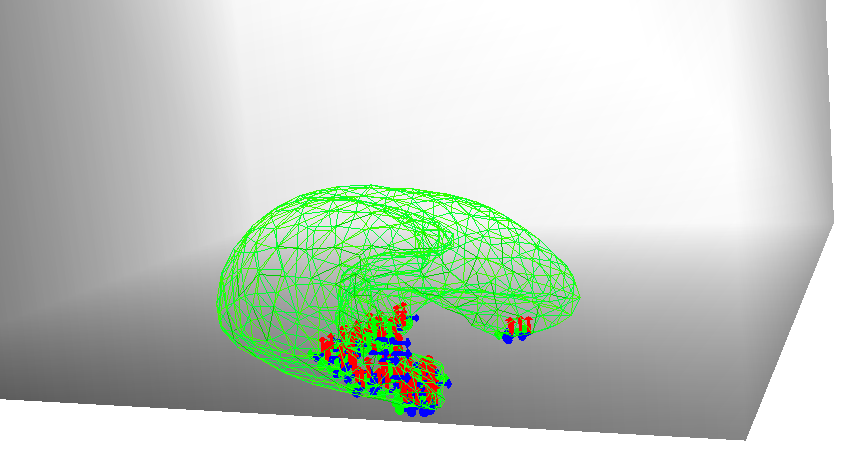}%
\caption{Liver collision}%
\label{fig:contactsimuliver}
\end{subfigure}%
\begin{subfigure}[t]{0.3\linewidth}%
\includegraphics[width=.9\linewidth]{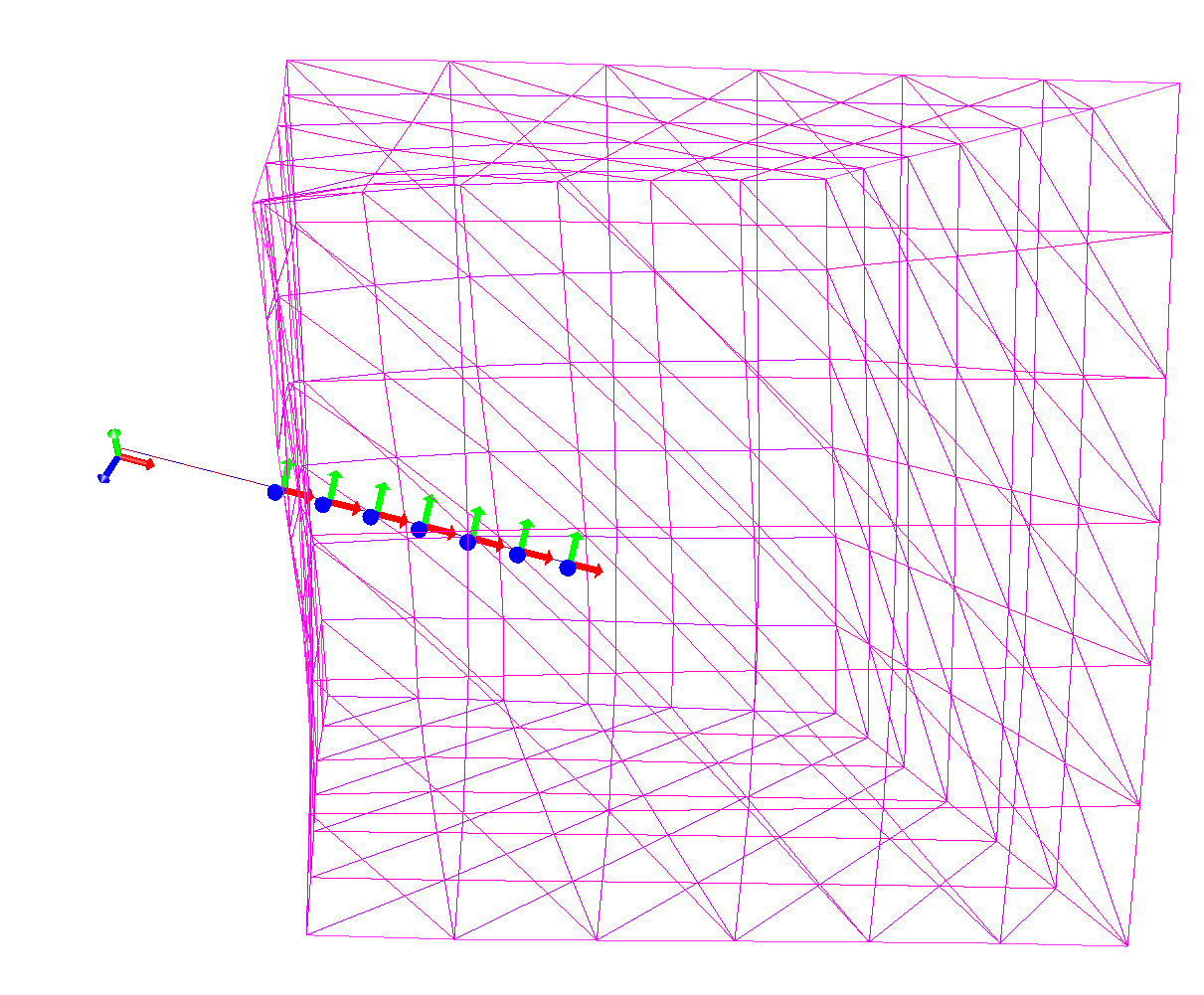}%
\caption{Needle insertion (cube)}%
\label{fig:contactsimuneedle}
\end{subfigure}%
\begin{subfigure}[t]{0.3\linewidth}%
\includegraphics[width=0.9\linewidth]{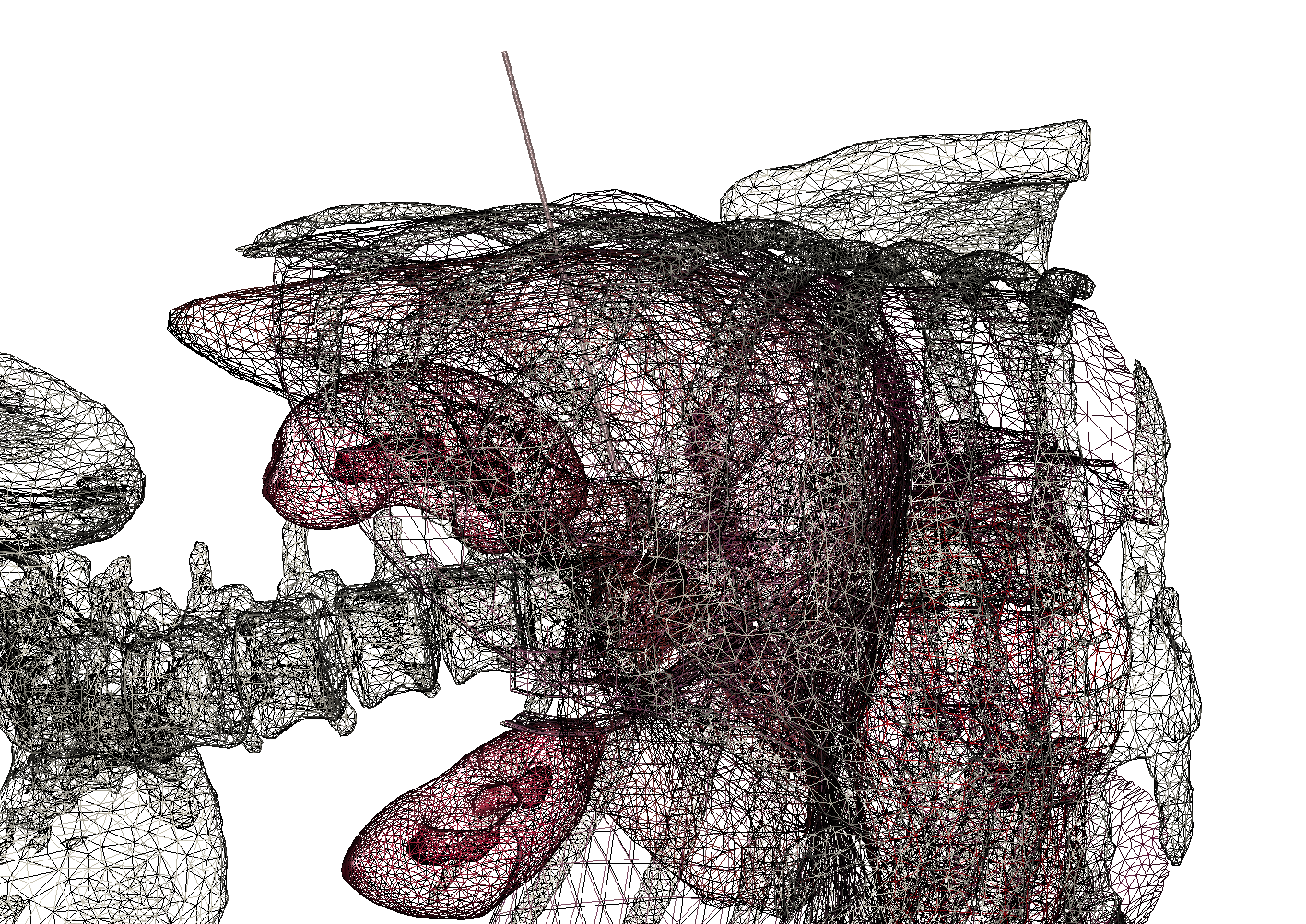}%
\caption{Needle insertion in multi-physics simulation.}%
\label{fig:contacthumanneedle}
\end{subfigure}%
\caption{Simulation including contacts and interactions.}%
\label{fig:contact simu}
\end{figure}

The fast assembly method is applied in various simulations with interactions such as simulating an object falling on rigid planes and needle insertion in soft bodies.
Interactions are based on Lagrange multipliers.
The compliance matrix $\textbf{W}$ is built using the asynchronous preconditioner, and constraints are solved using the GPU-based solver introduced in \cite{courtecuisse2014real}.
However, these simulations still benefit from the fast assembly method for the \textit{free motion} and \textit{motion correction}.

According to different scenarios, the speedup for the whole simulation depends on the ratio between the computation time for the \textit{constraint resolution} and the rest of the simulation.
For example, the simulation of the colliding deformable liver (see Figure \ref{fig:contactsimuliver}) involves $246$ constraints per step on average.
The \textit{constraint resolution} stage takes $56.7 \%$ of the all computation time of a single time step while the \textit{free motion} represents only $22.5 \%$.

Although the benefits of the fast assembly method are provided only when the fill ordering of the matrix remains constant, complex interaction can still be simulated. 
Figure \ref{fig:contactsimuneedle} shows a complex needle insertion in a soft body with complex interaction involving friction. 
Lagrangian multipliers are dynamically added to constrain the relative displacement of the needle and the cube when the needle penetrates the soft structure.
Despite the dynamic nature of the scene, the computation of the mapping is only performed at the initial step.
The needle insertion test into human organs is simulated in a heterogeneous scenario where the liver presents a co-rotational model while the skin covering the liver is modeled with Mooney-Rivlin material (see \ref{fig:contacthumanneedle}). 
The diaphragm and the intestine are modeled with hexahedral elements, whereas the liver and the skin are composed of tetrahedra. 
Therefore, the method proves its compatibility with the contact problem and significant flexibility to different elements and materials.
For these scenarios with complex interactions, similar speedup as shown previously in Table \ref{tab:precond} is observed during the \textit{free motion}.

\section{Conclusion}

This paper introduces a framework for real-time finite element simulations.
Besides its efficiency, our method remains generic for different constitutive models.
We propose a new matrix assembly strategy, which gains a significant speedup when the topology structure remains invariant and keeps the building cost on the same level as standard methods when the matrix pattern needs to be rebuilt.
The fast matrix assembly gives a possibility for parallelizations in the solving stage without any specific parallel implementation of the constitutive model.
Moreover, we replace the CPU-based preconditioner with a new GPU-based implementation in the solving stage.
This improvement significantly reduces the data transfer between CPU/GPU and makes running a fully GPU-based CG solver possible.
Finally, we evaluate our matrix assembly and parallelization strategy in various examples, including different element types and constitutive models.
Our approach is also proven to be compatible with contact problems.
We hope that our work will help researchers and engineers improve the performance of their works on FE simulations.




\bibliographystyle{siamplain}
\bibliography{mimesis}
\end{document}